\documentclass[twocolumn,aps,pra,superscriptaddress]{revtex4}
\usepackage[latin9]{inputenc}
\usepackage{amsmath}
\usepackage{amssymb}
\usepackage{graphicx}
\usepackage{bm}
\usepackage{color}
\usepackage{braket}
\usepackage[colorlinks=true,linkcolor=blue,citecolor=blue,urlcolor=blue]{hyperref}
\usepackage{subfigure}\usepackage{epsfig}\usepackage{amsfonts}\usepackage{mathrsfs}\usepackage[toc,page,title,titletoc,header]{appendix}
\makeatletter

\@ifundefined{textcolor}{}
{%
 \definecolor{BLACK}{gray}{0}
 \definecolor{WHITE}{gray}{1}
 \definecolor{RED}{rgb}{1,0,0}
 \definecolor{GREEN}{rgb}{0,1,0}
 \definecolor{BLUE}{rgb}{0,0,1}
 \definecolor{CYAN}{cmyk}{1,0,0,0}
 \definecolor{MAGENTA}{cmyk}{0,1,0,0}
 \definecolor{YELLOW}{cmyk}{0,0,1,0}
}

\makeatother

\begin{document}

\title{Control of Spin-Exchange Interaction between Alkali-Earth Atoms via Confinement-Induced Resonances
in a Quasi 1+0 Dimensional System}

\author{Ren Zhang}
\email{rine.zhang@gmail.com}

\affiliation{Department of Applied Physics, School of Science, Xi'an Jiaotong University, Xi'an, 710049, China}

\affiliation{Institute for Advanced Study, Tsinghua University, Beijing, 100084,
China}

\author{Peng Zhang}
\email{pengzhang@ruc.edu.cn}

\affiliation{Department of Physics, Renmin University of China, Beijing, 100872,
China}

\affiliation{Beijing Computational Science Research Center, Beijing, 100084, China}

\date{\today}
\begin{abstract}
A nuclear-spin exchange interaction exists between two ultracold fermionic alkali-earth (like) atoms in the electronic $^{1}{\rm S}_{0}$ state ($g$-state) and $^{3}{\rm P}_{0}$ state ($e$-state), and is an essential ingredient for the quantum simulation of Kondo effect. We study the control of this spin-exchange interaction for two atoms simultaneously confined in a quasi-one-dimensional (quasi-1D) tube, where the $g$-atom is freely moving in the axial direction while the $e$-atom is further localized by an additional axial trap and behaves as a quasi-zero-dimensional (quasi-0D) impurity. 
In this system, the two atoms experience effective-1D spin-exchange interactions in both even and odd partial wave channels, whose intensities can be controlled by the characteristic lengths of the confinements via the confinement-induced-resonances (CIRs).
In a previous work, we and our collaborators have studied this problem with a simplified pure-1D model (Phys. Rev. A {\bf 96}, 063605 (2017)). In current work, we go beyond that pure-1D approximation. We model the transverse and axial confinements by harmonic traps with finite characteristic lengths $a_\perp$ and $a_z$, respectively,  and 
 exactly solve the ``quasi-1D + quasi-0D" scattering problem between these two atoms. Using the solutions we derive the effective 1D spin-exchange interaction and investigate the locations and widths of the even/odd wave CIRs for our system. 
 It is found that when the ratio $a_z/a_\perp$ is larger, the CIRs can be induced by weaker confinements, which are easier to be realized experimentally.
 The comparison between our results and the recent experiment by  L. Riegger {\it et.al.} (Phys. Rev. Lett. {\bf 120}, 143601 (2018)) shows that the two experimentally observed resonance branches of the spin-exchange effect are due to an even-wave CIR and an odd-wave CIR, respectively. Our results are advantageous for the control and description of either the effective spin-exchange interaction or other types of interactions between ultracold atoms in quasi 1+0 dimensional systems.
\end{abstract}
\maketitle

\section{Introduction}

In recent years the ultracold gases of alkali-earth (like) atoms, e.g., Ca, Sr
and Yb, have attracted many attentions \cite{ca,review,hubbar1,hubbard2,hubbard3,martin,Jun,1dsun,ourprl,ofrexp1,ofrexp2,soc1,soc2,soc3,Munich,Florence}. One important application
of this system is the quantum simulation for the Kondo effect \cite{kondo} which
is induced by the spin-exchange between localized impurities and itinerant
fermions \cite{kondo-Rey,kondo-rey2,kondo-Jo,our1,our2,blochexp,kondo-salomon}. The following two features of alkali-earth (like)  atoms play a critical role in this quantum simulation: \begin{itemize}

\item[(i)] An alkali-earth (like) atom has not only a stable electronic orbital
ground state, i.e., the $^{1}{\rm S}_{0}$ state ($g$-state), but
also a very long-lived electronic orbital excited state, i.e., the
$^{3}{\rm P}_{0}$ state  ($e$-state), (Fig.~1(a)). These two states have different
AC polarizabilities except for the lasers with a magic wavelength \cite{magic1,magic2}.
Therefore, in experiments, one can realize either same or different
trapping potentials for the atom in $g$-state and $e$-state. 

\item[(ii)] There exists a spin-exchange interaction between two homonuclear fermionic alkali-earth (like) atoms in  $e$- and $g$-state. As a result,
these two atoms can exchange their nuclear-spin states during collision,
i.e., the process
\begin{equation}
|e,\uparrow\rangle|g,\downarrow\rangle\leftrightharpoons|e,\downarrow\rangle|g,\uparrow\rangle\label{se}
\end{equation}
can occur. 
\end{itemize}
Benefiting from these two properties, one can simulate the Kondo effect with ultracold alkali-earth (like) atoms in
an optical lattice which is very deep for the atoms in the $e$-state ($e$-atoms) and very
shallow for the atoms in the $g$-state ($g$-atoms). In that system the $e$-atoms are localized as
impurities and the $g$-atoms remain itinerant (Fig.~1(b)).

Nevertheless, to perform this quantum simulation
one still requires to enhance the intensity of the spin-exchange interaction
between the $g$-atom and $e$-atom, so that the Kondo temperature
can be high enough and thus attainable by current cooling capability.
In previous works \cite{our1,our2}, 
we and our collaborators proposed to solve this problem by confinement-induced
resonance (CIR). As shown in Fig. 1(c), in this scheme both the $g$-atoms
and the $e$-atoms are confined in a quasi-one-dimensional (quasi-1D)
confinement with the same characteristic length $a_{\perp}$, which
is generated by laser beams with magic wavelength. In addition, there
is also an confinement along the axial direction of the quasi-1D tube,
which can only be experienced by the $e$-atoms and has characteristic
length $a_{z}$. One can tune $a_\perp$ and $a_z$ by changing the intensities of the optical lattices.
As a result, the $g$-atoms are freely moving in the quasi-1D
tube, while the $e$-atoms are localized as quasi-zero-dimensional
(quasi-0D) impurities. Here we assume that in each axial confinement there
is one $e$-atom. In this system the $g$-atom and $e$-atom experience
an effective 1D spin-exchange interaction. The strength $\Omega$ of this effective interaction
is
determined by the scattering amplitude between these two atoms, which is a function of $a_\perp$ and $a_z$. CIR is the scattering resonance which occurs when $a_\perp$ and $a_z$ are tuned to some specific values.
At a CIR point $\Omega$
can be resonantly enhanced. 
In addition, when the system is near a CIR, one can efficiently control $\Omega$ by tuning
$a_{z}$ and $a_{\perp}$. 
\begin{figure}[t]
\begin{center}
\includegraphics[width=3.in]{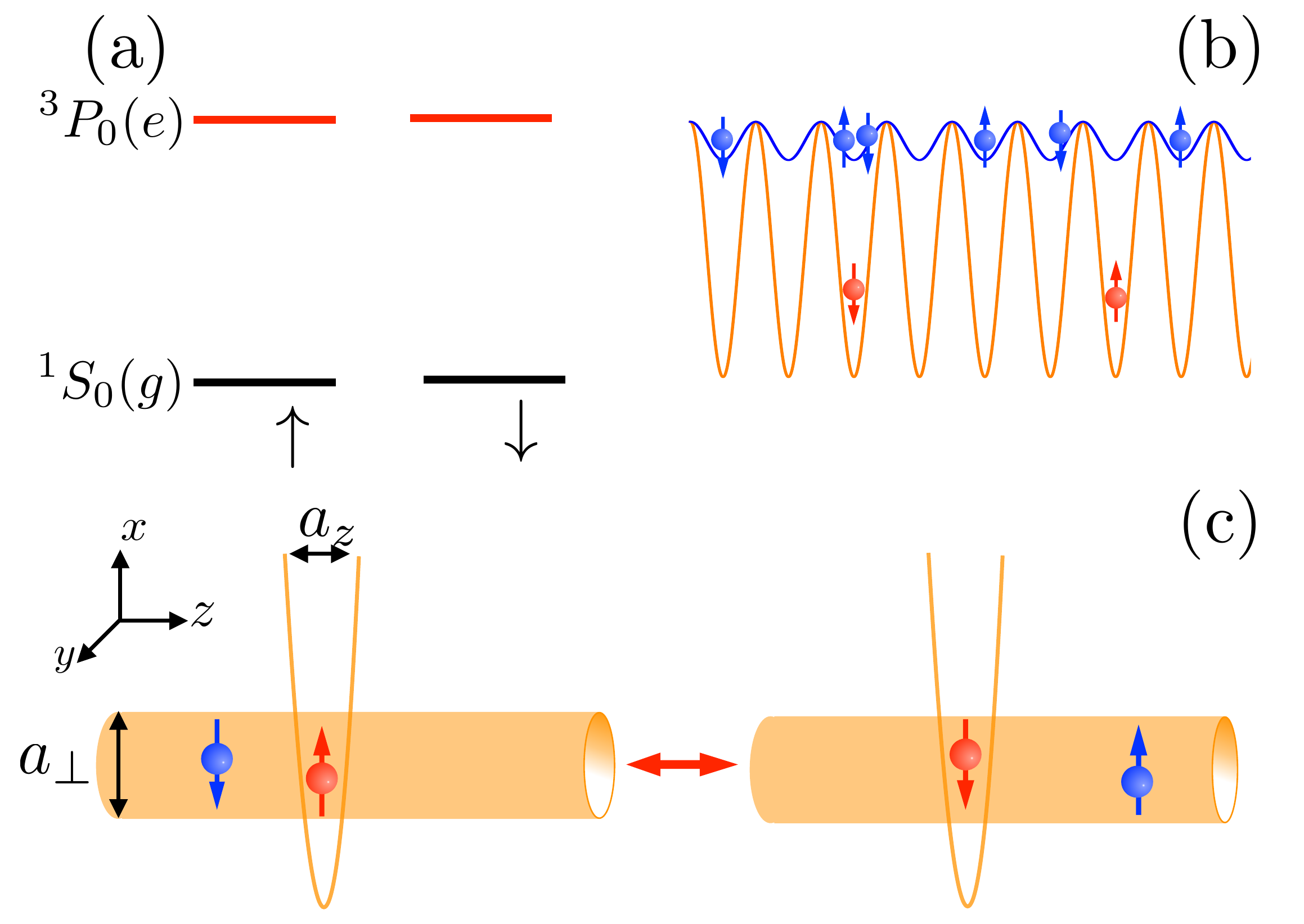}
\end{center}
\caption{(color online) {\bf (a):} Energy levels of a fermionic alkali-earth (like) atom. No matter if the atom is in the electronic orbit $^1{\rm S}_0$ state ($g$-state) or the $^3{\rm P}_0$ state ($e$-state), the nuclear spin could always be either $\uparrow$ or $\downarrow$.
{\bf (b):} Deep lattice for $e$-atoms  weak lattice for $g$-atoms created by the same standing-wave laser.  {\bf (c):} 
The quasi 1+0 dimensional system for the quantum simulation of Kondo effect.
The spin-exchange process occurs during the scattering between $g$-atoms (itinerant fermions) and localized $e$-atom (impurity).
In (b) and (c) the blue and red ball denote the $g$-atom and $e$-atom, respectively, and the arrows denote the nuclear spins.\label{schematic}}
\end{figure}

Our results in Refs. \cite{our1,our2} are
qualitatively consist with the recent experiment by L. Riegger {\it et.al}  \cite{blochexp},
where the quasi 1+0 dimensional system is realized with ultracold $^{173}$Yb
atoms, and the resonant control of the spin-exchange
strength via the CIRs is demonstrated.

On the other hand, for simplicity, some approximations are implemented in our works in Refs. \cite{our1} and \cite{our2}. In Ref. \cite{our1} we investigate the control of the effective-1D spin-exchange interaction strength  by tuning the quasi-1D confinement (transverse
confinement). Thus, we approximate  this strength
as the one for the systems where all atoms
are freely moving in the quasi-1D tube, i.e., the axial trap is ignored
in our two-body calculations. In Refs. \cite{our2} we focus on the effect induced by the axial trap for the $e$-atom. Accordingly, we ignore the transverse
degree of freedom and use a pure-1D model which only describes the
axial motion. These approximations are reasonable for the cases where
the characteristic lengths of the transverse and axial confinements,
i.e., $a_{\perp}$ and $a_{z}$, are very different from each other.

However, in realistic systems   $a_{\perp}$
and $a_{z}$ are generally of the same order. In this case, the cross effect
of the transverse and axial confinements can be important. Thus, we
should go beyond the above approximations and explicitly take into
account both of the two confinements into the theoretical calculation.

In this paper, we perform such a complete calculation. In our model the
transverse and axial confinements are described by harmonic potentials
with finite characteristic lengths $a_{\perp}$ and $a_{z}$, respectively.
We exactly solved the ``quasi-1D
+ quasi-0D" scattering problem  between a freely-moving
$g$-atom and a trapped $e$-atom.
In this system, there are two partial-wave scattering channels, i.e., the even-wave channel and the odd-wave channels. 
We derive the scattering amplitude for both of these two partial waves, as well as the effective 1D interaction between these two atoms. 
Using these results we investigate the even- and odd-wave CIRs in our systems. It is found that when the ratio $a_z/a_\perp$ is larger, 
the CIRs can occur in weaker confinements (i.e., the confinements with lower trapping frequencies), which are easier to be realized in experiments.

We further compare our results with the recent experimental observations shown in Ref. \cite{blochexp}. The authors of Ref. \cite{blochexp} have done a theoretical calculation based on a two-site model, where
the coupling between the center-of-mass motion and relative motion of two atoms in the same site is ignored.
Here we use our model to explore the location and widths of the CIRs in this experiment. 
As shown above, in our model the optical-lattice-induced confinement potentials are approximated as harmonic potentials. 
In addition, in the experiments, the $g$-atoms also experience a shallow lattice potential in the axial direction, 
and in our current calculation, we ignore this shallow lattice.   Nevertheless, even with these two simplifications, our results are still
consistent well with the experiment.  In particular, our results reveal that one resonance branch of the  spin-exchange effect observed  in the experiment is due to an even-wave CIR, while the other one is due to an odd-wave CIR. Thus, the effective spin-exchange interaction for these two resonance branches are quite different with each other.

Our results are valuable
for the quantum simulation of Kondo effect with alkali-earth (like) atoms. Furthermore, our exact solution for the ``quasi-1D
+ quasi-0D" scattering problem can also be applied to other problems of quasi-1D ultracold gases with localized impurities, e.g., the realization of high-precision magnetometer with such a system \cite{magton,magton2}.

The remainder of this paper is organized as follows. In Sec. II we
describe the detail of our model and the approach of our calculation.
In Sec. III we illustrate our results and investigate the CIR effects
for our system. In this section, we also compare our results with the
experimental results. A summary and discussion
are given in Sec. IV. In the appendix we present details of our calculation.

\section{Effective 1D Interaction}

\subsection{Model}

As shown above, we consider two ultracold fermionic alkali-earth (like) atoms of the same species,
with one atom being in the electronic $g$-state and the other one being in the $e$-state. In our problem
the electronic states $e$ and $g$ can be used as the labels of the two atoms, i.e., the  $e$-atom and $g$-atom behave as two distinguishable particles. Accordingly, 
the nuclear-spin states of the $g$-  ($e$-) atom-atom can be denoted as $|\uparrow\rangle_{g(e)}$ and $|\downarrow\rangle_{g(e)}$.
Here we consider the case with zero magnetic fields, i.e., $B=0$. We 
 assume the two atoms are tightly confined in a two-dimensional
isotropic harmonic trap in the $x-y$ plane (Fig 1(c)), which is formed
by laser beams with magic wavelength and thus has the same intensity
for both of the two atoms. In addition, there is also an axial harmonic
trap in the $z$-direction, which is only experienced by the $e$-atom.
We further define ${\bf r}\equiv(x_{r},y_{r},z_{r})$ as the relative
coordinate of these two atoms, and $z_{g(e)}$ as the $z$-coordinate
of the $g$- ($e$-) atom which satisfy $z_{r}=z_{g}-z_{e}$. 

The Hamiltonian for the two-body problem is given by
\begin{equation}
H=H_{0}+V,\label{h}
\end{equation}
with $H_{0}$ and $V$ being the free Hamiltonian and the inter-atomic
interaction in three-dimensional (3D) space, respectively. Furthermore,
in the $x-y$ plane the relative motion of the two atoms can be decoupled
from the center-of-mass motion. Therefore, the free Hamiltonian $H_{0}$
can be expressed as ($m=\hbar=1$, with $m$ being the single-atom
mass)

\begin{equation}
H_{0}=-\frac{1}{2}\frac{\partial^{2}}{\partial z_{g}^{2}}+H_{\perp}+H_{e},\label{h0}
\end{equation}
with
\begin{eqnarray}
H_{\perp} & = & -\frac{\partial^{2}}{\partial x_{r}^{2}}-\frac{\partial^{2}}{\partial y_{r}^{2}}+\frac{\omega_{\perp}^{2}}{4}(x_{r}^{2}+y_{r}^{2});\label{hperp}\\
H_{e} & = & -\frac{1}{2}\frac{\partial^{2}}{\partial z_{e}^{2}}+\frac{\omega_{z}^{2}}{2}z_{e}^{2},\label{he}
\end{eqnarray}
where $\omega_{\perp}$ and $\omega_{z}$ are the frequency of the
transverse and axial confinements, respectively. They are related to the characteristic lengths $a_\perp$ and $a_z$ via
\begin{eqnarray}
a_{\perp} =\sqrt{\frac{2}{\omega_\perp}};\ a_{z} =\sqrt{\frac{1}{\omega_z}}.
\end{eqnarray}

In addition, for our system the inter-atomic interaction $V$ is diagonal
in the basis of nuclear-spin singlet and triplet states:
\begin{eqnarray}
|+\rangle & = & \frac{1}{\sqrt{2}}\left(|\uparrow\rangle_{g}|\downarrow\rangle_{e}-|\downarrow\rangle_{g}|\uparrow\rangle_{e}\right);\label{p}\\
|-,0\rangle & = & \frac{1}{\sqrt{2}}\left(|\uparrow\rangle_{g}|\downarrow\rangle_{e}+|\downarrow\rangle_{g}|\uparrow\rangle_{e}\right);\label{m0}\\
|-,+1\rangle & = & |\uparrow\rangle_{g}|\uparrow\rangle_{e};\label{m1}\\
|-,-1\rangle & = & |\downarrow\rangle_{g}|\downarrow\rangle_{e},\label{mn1}
\end{eqnarray}
and can be expressed as
\begin{equation}
V=V_{+}{\cal P}_++V_{-}{\cal P}_-,\label{vv}
\end{equation}
where 
\begin{eqnarray}
{\cal P}_+=|+\rangle\langle+|,\ \ {\cal P}_-=\sum_{q=0,\pm1}|-,q\rangle\langle-,q|.
\end{eqnarray}
Here $V_{+}$ and $V_{-}$ are the interaction potential in the channels
of nuclear-spin singlet and triplet state, respectively. They can
be modeled by Huang-Yang pseudo potential
\begin{equation}
V_{\pm}=4\pi a_{\pm}\delta({\bf r})\frac{\partial}{\partial r}\left(r\cdot\right),\label{hy}
\end{equation}
with $r=|{\bf r}|$, and $a_{\pm}$ are the corresponding $s$-wave
scattering lengths. For a certain type of alkali-earth (like) atom, the two scattering
lengths $a_{+}$ and $a_{-}$ are usually different. For instance,
for $^{173}$Yb atoms we have $a_{+}\approx1878a_{0}$ and $a_{-}\approx216a_{0}$,
with $a_{0}$ being the Bohr's radius \cite{ofrexp2}. On the other hand, Eq. (\ref{vv})
directly yields that $_{g}\!\langle\downarrow|{}_{e}\!\langle\uparrow|V|\uparrow\rangle_{g}|\downarrow\rangle_{e}\propto(a_{+}-a_{-})$.
Thus, the strength of the spin-exchange interaction in 3D space is
proportional to $(a_{+}-a_{-})$.

In this work, we consider the cases that the temperature is much lower
than $\omega_{\perp}/k_{B}$ and $\omega_{z}/k_{B} $, with $k_{B}$
being the Boltzmann constant. In these cases, when the two atoms are
far away from each other, the relative motion in the $x-y$ plane
and the axial motion of the $e$-atom in the $z$-direction are frozen in
the ground states of the corresponding harmonic confinements. As a
result, our system can be effectively described by a simple model where the $g$-atom and $e$-atom are spin-1/2 particles, which are freely moving in the pure-1D space and fixed at $z=0$, respectively.
The effective Hamiltonian
of this pure 1D model can be expressed as
{
\begin{equation}
H_{{\rm eff}}=-\frac{1}{2}\frac{\partial^{2}}{\partial z_{g}^{2}}+V_+^{\rm (eff)}{\cal P}_++
V_-^{\rm (eff)}{\cal P}_-,\label{heff}
\end{equation}
where $V_{+/-}^{\rm(eff)}$ is the effective potential for the nuclear-spin singlet/triplet states. In the 1D scattering problem between the freely-moving $g$-atom and the fixed $e$-atom, there are two partial-wave scattering channels, i.e., the even wave and the odd wave. As a result, $V_{\xi}^{\rm(eff)}$ ($\xi=+,-$) can be expressed as the summation of the 1D zero-range pseudo potentials for these two partial waves. Explicitly, we have \cite{olshaniiodd}
\begin{equation}
V_{\xi}^{\rm(eff)}=g_{\xi}^{\rm(even)}\delta(z_g){\hat d}_{\rm e}+g_{\xi}^{\rm(odd)}\delta^\prime(z_g){\hat d}_{\rm o},\ \ ({\rm for}\ \xi=+,-).\label{veff}
\end{equation}
Here  $\delta(z_g)$ is the Dirac delta function, $\delta^\prime(z_g)=\frac{d\delta(z_g)}{dz_g}$ and the operators ${\hat d}_{e}$ and ${\hat d}_{o}$ are defined as
\begin{eqnarray}
{\hat d}_{\rm e}\psi(z_g)&\equiv&\frac{1}{2}\left[\left.\psi(z_g)\right|_{z_g=0^+}+\left.\psi(z_g)\right|_{z_g=0^-}\right],\nonumber\\
\label{de}\\
{\hat d}_{\rm o}\psi(z_g)&\equiv&\frac{1}{2}\left[\left.\frac{d}{dz_g}\psi(z_g)\right|_{z_g=0^+}+\left.\frac{d}{dz_g}\psi(z_g)\right|_{z_g=0^-}\right].\nonumber\\
\label{do}
\end{eqnarray}
The operators ${\hat d}_{\rm e}$ and ${\hat d}_{\rm o}$ are essentially the projection operators to the even and odd partial wave channels, respectively, and $g_{\xi}^{\rm(even)}\delta(z_g){\hat d}_{\rm e}$ and $g_{\xi}^{\rm(odd)}\delta^\prime(z_g){\hat d}_{\rm o}$ are the 1D even- and odd-wave pseudo potentials, respectively  \cite{olshaniiodd}.

Furthermore, the complete effective potential $V_+^{\rm (eff)}{\cal P}_++
V_-^{\rm (eff)}{\cal P}_-$ is required to reproduce the correct low-energy scattering
amplitude between the freely-moving $g$-atom and the $e$-atom in the ground
state of the axial trap. Comparing Eqs. (\ref{heff}, \ref{veff}) with Eqs.
(\ref{h}, \ref{vv}), one can find that this means that the low-energy
scattering amplitude for the Hamiltonian $-\frac{1}{2}\frac{\partial^{2}}{\partial z_{g}^{2}}+g_{\xi}^{\rm(even)}\delta(z_g){\hat d}_{\rm e}+g_{\xi}^{\rm(odd)}\delta^\prime(z_g){\hat d}_{\rm o}$
should approximately equal to the one for $H_{0}+V_{\xi}$ ($\xi=+,-$).
This requirement determines the value of the intensities $g_{\pm}^{({\rm even})}$ and $g_{\pm}^{({\rm odd})}$. 

On the other hand, the complete effective interaction $V_+^{\rm (eff)}{\cal P}_++
V_-^{\rm (eff)}{\cal P}_-$ can be re-written
as 
\begin{eqnarray}
&&V_+^{\rm (eff)}{\cal P}_++
V_-^{\rm (eff)}{\cal P}_-\nonumber\\
&=&\Lambda(z_{g})
 +\Omega(z_{g})\left[\frac12\sigma_{z}^{(g)}\sigma_{z}^{(e)}+\sigma_{+}^{(g)}\sigma_{-}^{(e)}+\sigma_{-}^{(g)}\sigma_{+}^{(e)}\right],\nonumber\\
 \label{veff2}
\end{eqnarray}
where $\sigma_{z}^{(j)}=|\uparrow\rangle_{j}\langle\uparrow|-|\downarrow\rangle_{j}\langle\downarrow|$,
$\sigma_{+}^{(j)}=|\uparrow\rangle_{j}\langle\downarrow|$ and $\sigma_{-}^{(j)}=\sigma_{+}^{(j)\dagger}$ ($j=e,g$ )
are the Pauli operators for the $j$-atom, and the $\Lambda(z_{g})$ and $\Omega(z_g)$
are defined as
\begin{eqnarray}
\Lambda(z_g)&=&\Lambda^{\rm (even)}\delta(z_g){\hat d}_{\rm e}+\Lambda^{\rm (odd)}\delta^\prime(z_g){\hat d}_{\rm o};\\
\Omega(z_g)&=&\Omega^{\rm (even)}\delta(z_g){\hat d}_{\rm e}+\Omega^{\rm (odd)}\delta^\prime(z_g){\hat d}_{\rm o},\label{omed}
 \end{eqnarray} 
with
\begin{eqnarray}
\Lambda^{\rm (even/odd)}&=&\frac34g_{-}^{{\rm (even/odd)}}+\frac14g_{+}^{{\rm (even/odd)}};\label{lam}\\
 \Omega^{\rm (even/odd)}&=&\frac12\left[g_{-}^{({\rm even/odd})}-g_{+}^{({\rm even/odd})}\right].\label{ome}
 \end{eqnarray} 
Thus, $\Omega^{\rm (even)}$ and $\Omega^{\rm (odd)}$ indicate the strenght of the  effective 1D spin-exchange interaction.

Since the 1D effective interaction $V_+^{\rm (eff)}{\cal P}_++
V_-^{\rm (eff)}{\cal P}_-$ is determined by the four parameters $g_{\pm}^{({\rm even/odd})}$, in the next subsection we calculate  $g_{\pm}^{({\rm even/odd})}$ 
via solving the two-atom scattering problem.
}

\subsection{``Quasi-1D + Quasi-0D" Scattering Problem}

As shown
above, the value of $g_{\xi}^{({\rm even/odd})}$ ($\xi=+,-$) is determined
by the scattering amplitude between a $g$-atom moving in the quasi-1D
confinement and an $e$-atom localized by the axial trap, with two-atom Hamiltonian $H_{0}+V_{\xi}$.
Thus, to calculate $g_{\xi}^{({\rm even/odd})}$ we first solve this ``quasi-1D + quasi-0D" scattering
problem. Our approach is similar
to that of P. Massignan and Y. Castin \cite{castin}, who calculated
the scattering amplitude between one atom freely moving in 3D free space
and another atom localized in a 3D harmonic trap.

{\subsubsection{Scattering amplitudes}}

In the incident state of our problem, the relative transverse motion
of the two atoms and the axial motion of the $e$-atom are in the 
ground states of the corresponding confinements. Therefore, the incident wave function $\Psi^{(0)}(\bm{\rho},z_{e},z_{g})$
can be expressed as
\begin{equation}
\Psi^{(0)}(\bm{\rho},z_{e},z_{g})=\frac{e^{ikz_{g}}}{\sqrt{2\pi}}\chi_{n_{\perp}=0,m_{z}=0}(\bm{\rho})\phi_{n_{z}=0}(z_{e}),\label{psik0}
\end{equation}
where $k$ is the incident momentum of the $g$-atom, and $\bm{\rho}=x_{r}{\bf e}_{x}+y_{r}{\bf e}_{y}$
is the transverse relative position vector, with ${\bf e}_{x(y)}$
being the unit vector along the $x$- ($y$-) direction. Here $\chi_{n_{\perp},m_{z}}(\bm{\rho})$
is the eigen-state of the transverse relative Hamiltonian $H_{\perp}$
defined in Eq. (\ref{hperp}), with $n_{\perp}$ and $m_{z}$ being
the principle quantum number and the quantum number of the angular
momentum along the $z$-direction. It satisfies 
\begin{equation}
H_{\perp}\chi_{n_{\perp},m_{z}}(\bm{\rho})=(n_{\perp}+1)\omega_{\perp}\chi_{n_{\perp},m_{z}}(\bm{\rho}),\label{hperp2}
\end{equation}
with $m_{z}=0,\pm1,\pm2,....$ and $n_{\perp}=|m_{z}|,|m_{z}|+2,|m_{z}|+4,...$
. In addition, in Eq. (\ref{psik0}) the function $\phi_{n_{z}}(z_{e})$
($n_{z}=0,1,2,...$) is the eigen-state of the axial Hamiltonian $H_{e}$
of the $e$-atom and satisfies
\begin{equation}
H_{e}\phi_{n_{z}}(z_{e})=\left(n_{z}+\frac{1}{2}\right)\omega_{z}\phi_{n_{z}}(z_{e}).\label{he2}
\end{equation}
Actually this function can be expressed as  $\phi_{n_{z}}(z_{e})=\sqrt{1/(a_{z}\sqrt{\pi}2^{n_{z}}n_{z}!)}e^{-\frac{z_{e}^{2}}{2a_{z}^{2}}}H_{n_{z}}(z_{e}/a_{z})$,
with $H_{n_{z}}$ being the Hermit polynomial.
It is clear that $\Psi^{(0)}$ is an eigen-state of the total free
Hamiltonian $H_{0}$, with eigen-value 
\begin{equation}
E=\frac{k^{2}}{2}+\omega_{\perp}+\frac{\omega_{z}}{2}.\label{ee}
\end{equation}
Here we assume that the incident kinetic energy $k^{2}/2$ is smaller
than the energy gap between the ground and first excited state of $H_{e}$
or $H_{\perp}$ with $m_{z}=0$, i.e., 
\begin{equation}
0\leq\frac{k^{2}}{2}<2\omega_{\perp};\ 0\leq\frac{k^{2}}{2}<\omega_{z}.\label{enecon}
\end{equation}

The scattering wave function $\Psi_{\xi}(\bm{\rho},z_{e},z_{g})$ corresponding to the incident state $\Psi^{(0)}$
is determined by the Schr$\ddot{{\rm o}}$dinger equation $(H_{0}+V_{\xi})\Psi_{\xi}=E\Psi_{\xi}$, with $V_\xi$ ($\xi=+,-$) being given by Eq. (\ref{hy}),
as well as the out-going boundary condition in the limit $|z_{g}|\rightarrow\infty$. 
These requirements can be equivalently reformulated as the integral
equation \cite{castin}
\begin{eqnarray}
\Psi_{\xi}(\bm{\rho},z_{e},z_{g}) & = & \Psi^{(0)}(\bm{\rho},z_{e},z_{g})+\nonumber \\
 &  & 4\pi a_{\xi}\int dz'G_{E}(\bm{\rho},z_{e},z_{g};{\bf 0},z^{\prime},z')\eta_{\xi}(z'),\nonumber \\
\label{lse}
\end{eqnarray}
where the function $\eta_{\xi}(z')$ is the regularized scattering wave function and is defined as
\begin{equation}
\eta_{\xi}(z')=\left.\frac{\partial}{\partial z_{r}}\left[z_{r}\Psi_{\xi}\left({\bf 0},z'-\frac{z_{r}}{2},z'+\frac{z_{r}}{2}\right)\right]\right|_{z_{r}\rightarrow0^{+}},\label{etaz}
\end{equation}
and $G_{E}$ is the retarded Green\textquoteright s function for the
free Hamiltonian $H_{0}$. Using the Dirac bracket we can express $G_{E}$
as
\begin{widetext}
\begin{eqnarray}
 G_{E}(\bm{\rho},z_{e},z_{g};\bm{\rho^{\prime}},z_{e}^{\prime},z_{g}^{\prime})= \langle\bm{\rho},z_{e},z_{g}|\frac{1}{E+i0^{+}-H_{0}}|\bm{\rho^{\prime}},z_{e}^{\prime},z_{g}^{\prime}\rangle,\label{ge}
\end{eqnarray}
where $|\bm{\rho},z_{e},z_{g}\rangle$ and $|\bm{\rho^{\prime}},z_{e}^{\prime},z_{g}\rangle$
are the eigen-states of the transverse relative position and the axial
coordinates of the $g$- and $e$- atoms.

We can extract the scattering amplitude from the behavior of $\Psi_{\xi}(\bm{\bm{\rho}},z_{e},z_{g})$
in the long-range limit $|z_{g}|\rightarrow\infty$. To this end,
we re-express the Green's function $G_{E}(\bm{\rho},z_{e},z_{g};{\bf 0},z^{\prime},z')$
as
\begin{eqnarray}
G_{E}(\bm{\rho},z_{e},z_{g};{\bf 0},z^{\prime},z') & =-i & \frac{e^{ik|z_{g}-z^{\prime}|}}{k}\chi_{n_{\perp}=0,m_{z}=0}(\bm{\rho})\chi_{n_{\perp}=0,m_{z}=0}^{\ast}({\bf 0})\phi_{n_{z}=0}(z_{e})\phi_{n_{z}=0}^{*}(z^{\prime})\nonumber \\
 &  & -\sum_{\substack{
n_{z}=0,1,2,...;\\
n_{\perp}=0,2,4,...;\\
(n_{z},n_{\perp})\neq(0,0)
}}\frac{e^{-\kappa_{n_{\perp},n_{z}}|z_{g}-z^{\prime}|}}{\kappa_{n_{\perp},n_{z}}}\chi_{n_{\perp},m_{z}=0}(\bm{\rho})\chi_{n_{\perp},m_{z}=0}^{\ast}({\bf 0})\phi_{n_{z}}(z_{e})\phi_{n_{z}}^{*}(z^{\prime}),\label{ge2}
\end{eqnarray}
with $\kappa_{n_{\perp},n_{z}}=\sqrt{2[(n_{\perp}+1)\omega_\perp+(n_{z}+1/2)\omega_{z}]-k^{2}}$.
In the derviation of Eq. (\ref{ge2}) we have used the fact that $\chi_{n_{\perp},m_{z}}(\bm{0})=0$
for $m_{z}\neq0$. Furthermore, due to the low-energy assumption (\ref{enecon}),
in the limit $|z_{g}|\rightarrow\infty$ all the terms in the summation
in Eq. (\ref{ge2}) decay to zero. Substituting Eq. (\ref{ge2}) into
Eq. (\ref{lse}) and using this result, we obtain
\begin{eqnarray}
 \Psi_{\xi}(\bm{\rho},z_{e},|z_{g}|\rightarrow\infty)= \frac{1}{\sqrt{2\pi}}\left[e^{ikz_{g}}+f_{\xi}^{{\rm even}}(k)e^{ik|z_{g}|}+f_{\xi}^{{\rm odd}}(k){\rm sign}(z_{g})e^{ik|z_{g}|}\right] \chi_{n_{\perp}=0,m_{z}=0}(\bm{\rho})\phi_{n_{z}=0}(z_{e}).\label{psi2}
\end{eqnarray}
Here the scattering amplitudes $f_{\xi}^{{\rm even}}(k)$ and $f_{\xi}^{{\rm odd}}(k)$ can be expressed as \cite{our2}
\begin{eqnarray}
f_{\xi}^{\rm even/odd}(k) & = & -i\frac{2(2\pi)^{\frac{3}{2}}a_{\xi}}{k}\chi_{n_{\perp}=0,m_{z}=0}^{\ast}({\bf 0})\int dz^{\prime}F_{\rm even/odd}(k,z^{\prime})\phi_{n_{z}=0}^{*}(z^{\prime})\eta_{\xi}(z^{\prime}),\label{fxi}
\end{eqnarray}
with $F_{{\rm even}}(k,z^{\prime})=\cos(kz^{\prime})$ and $F_{{\rm odd}}(k,z^{\prime})=-i\sin(kz^{\prime})$. 
It is clear that Eq. (\ref{psi2}) can be re-expressed in a convenient form ($k>0$)
\begin{eqnarray}
 \Psi_{\xi}(\bm{\rho},z_{e},|z_{g}|\rightarrow\infty)= \chi_{n_{\perp}=0,m_{z}=0}(\bm{\rho})\phi_{n_{z}=0}(z_{e})\times\left\{
 \begin{array}{lll}
 \frac{1}{\sqrt{2\pi}}\left[e^{ikz_g}+r_\xi(k)e^{-ikz_g}\right]&&({\rm for}\ z_g\rightarrow-\infty)\\
  \frac{1}{\sqrt{2\pi}}t_\xi(k)e^{ikz_g}&&({\rm for}\ z_g\rightarrow+\infty)
 \end{array}
 \right.,\label{newpsi}
\end{eqnarray}
\end{widetext}
where $r_\xi(k)$ and $t_\xi(k)$ are the reflection and transmission amplitudes, respectively, and are related to $f_{\xi}^{{\rm even/odd}}(k)$ via
 \begin{eqnarray}
r_\xi(k)&=&f_\xi^{\rm even}(k)-f_\xi^{\rm odd}(k);\\
t_\xi(k)&=&f_\xi^{\rm even}(k)+f_\xi^{\rm odd}(k)+1.
\end{eqnarray}

Actually,  $f_{\xi}^{{\rm even/odd}}(k)$ are nothing but the two partial-wave scattering amplitudes. Explicitly,  the complete Hamiltonian $H$ in Eq. (\ref{h}) is invariable under the total  reflection operation 
\begin{eqnarray}
T:\{z_g\rightarrow-z_g,z_e\rightarrow-z_e\}.\label{reft}
\end{eqnarray} 
As a result, the parity ${\mathbb P}$ with respect to this  reflection operation  is conserved. Therefore,  there are two partial-waves for our ``quasi-1D + quasi-0D" scattering problem, i.e., the even-wave (corresponding to ${\mathbb P}=+1$) and the odd-wave (corresponding to ${\mathbb P}=-1$). As shown in Appendix A, it can be proved that $f_{\xi}^{{\rm even/odd}}(k)$ given by Eq. (\ref{fxi}) are just the scattering amplitudes for the even/odd partial waves, respectively.

Here we would like to emphasis that, even though
our 3D bare interaction $V_{\pm}$ defined in Eq. (\ref{hy}) only includes the 
 $s$-wave component,
both of the even- and odd-wave scattering amplitudes $f_{\xi}^{{\rm even/odd}}(k)$ are non-zero. This can be explained as follows. The total parity ${\mathbb P}$ with respect to the reflection $T$ can be expressed as
\begin{eqnarray}
{\mathbb P}={\mathbb P}_{\rm CoM}\times{\mathbb P}_r,
\end{eqnarray}
where ${\mathbb P}_{\rm CoM}$ is the parity corresponding to the reflection of the center-of-mass coordinate (i.e., the transformation $\{Z\rightarrow-Z;z_r\rightarrow z_r\}$, with $Z=(z_e+z_g)/2$ and $z_r=z_g-z_e$ as defined above), and ${\mathbb P}_{r}$ is the parity corresponding to the reflection of the relative coordinate (i.e., the transformation $\{Z\rightarrow Z;z_r\rightarrow -z_r\}$). Therefore, in the odd-wave subspace (i.e., the subspace with ${\mathbb P}=-1$), there are some states with   ${\mathbb P}_{\rm CoM}=-1$ and ${\mathbb P}_r=+1$. Thus, although the $s$-wave Huang-Yang pseudo potentials $V_{\pm}$  only operates   on the states with  ${\mathbb P}_r=1$, it has non-zero projection for the odd-wave subspace. As a result, the odd-wave scattering amplitude $f_{\xi}^{{\rm odd}}(k)$ is non-zero. Similarly,  $f_{\xi}^{{\rm even}}(k)$ is also non-zero. It has been shown that in the  scattering problems of two ultracold atoms in a mixed-dimensional system, even if the inter-atomic interaction is described by a $s$-wave Huang-Yang pseudo potential, the high partial wave scattering amplitudes are usually non-zero \cite{tannishida}.

\subsubsection{Calculation of $f_{\xi}^{{\rm even/odd}}(k)$ }

Eq. (\ref{fxi}) shows that the scattering amplitudes $f_{\xi}^{{\rm even/odd}}(k)$, are functionals
of the regularized wave function $\eta_{\xi}(z^{\prime})$ defined in Eq. (\ref{etaz}). 
On the other hand, substituting Eq. (\ref{lse}) into Eq. (\ref{etaz}),
we can find that $\eta_\xi(z)$ satisfies another integral
equation (Appendix B)
\begin{equation}
\eta_{\xi}(z)=\Psi^{(0)}\left({\bf 0},z,z\right)+\hat{O}_{\xi}[\eta_{\xi}(z)].\label{eta2}
\end{equation}
Here $\hat{O}_{\xi}$ is an integral operator with the explicit form
being given in Appendix B. 

\begin{figure*}
\begin{center}
\includegraphics[width=5.5cm]{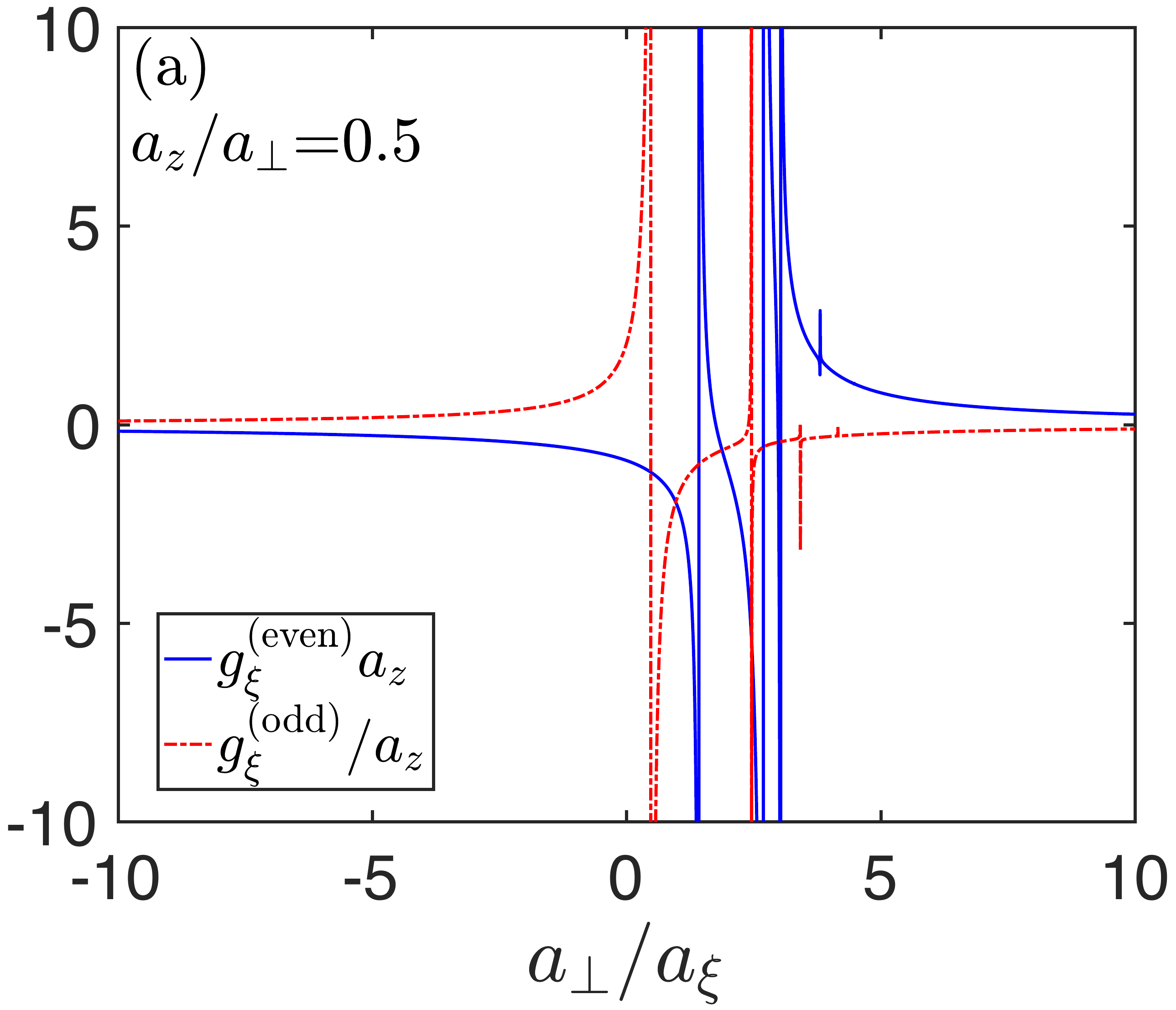}
\includegraphics[width=5.5cm]{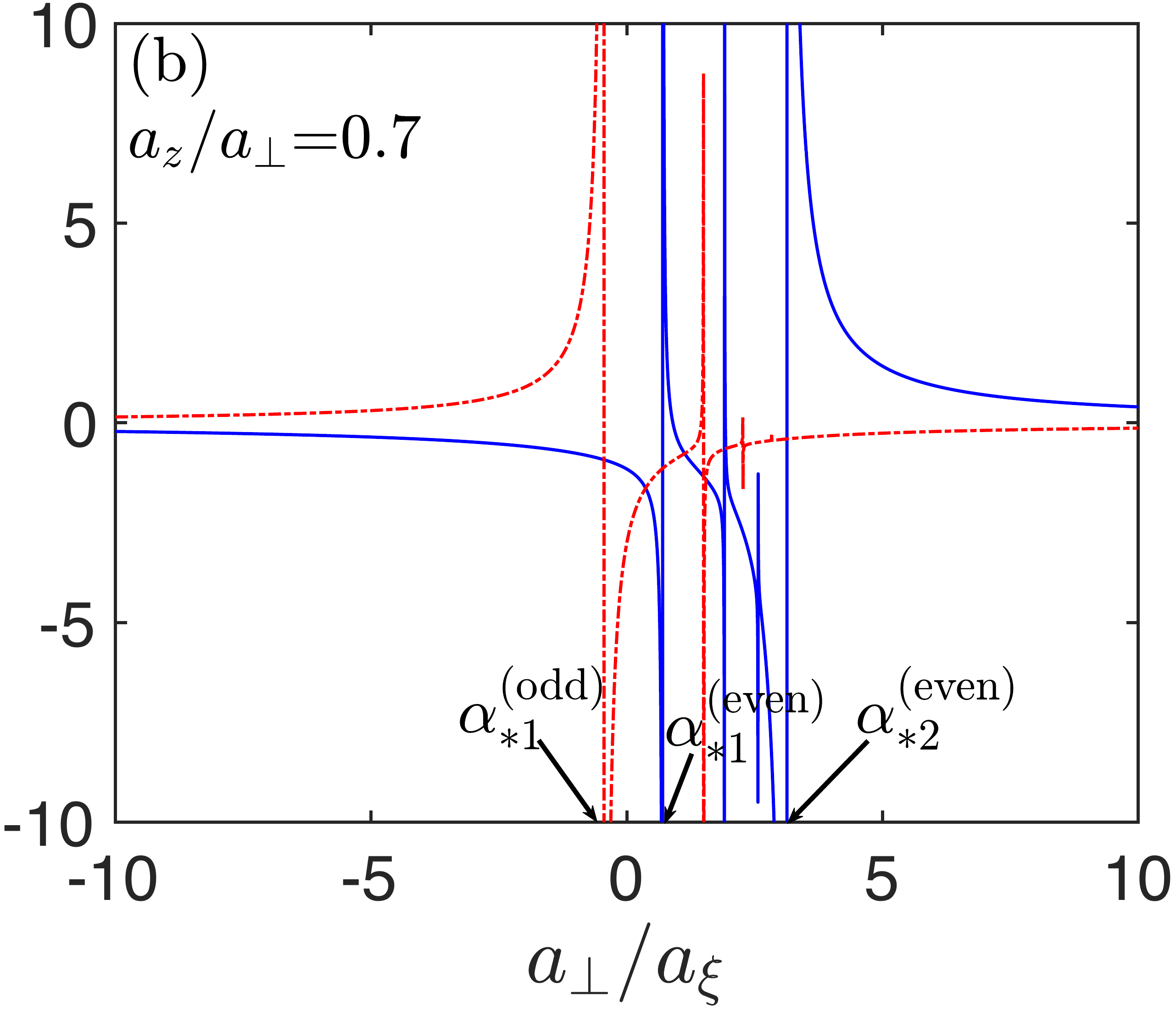}
\includegraphics[width=5.5cm]{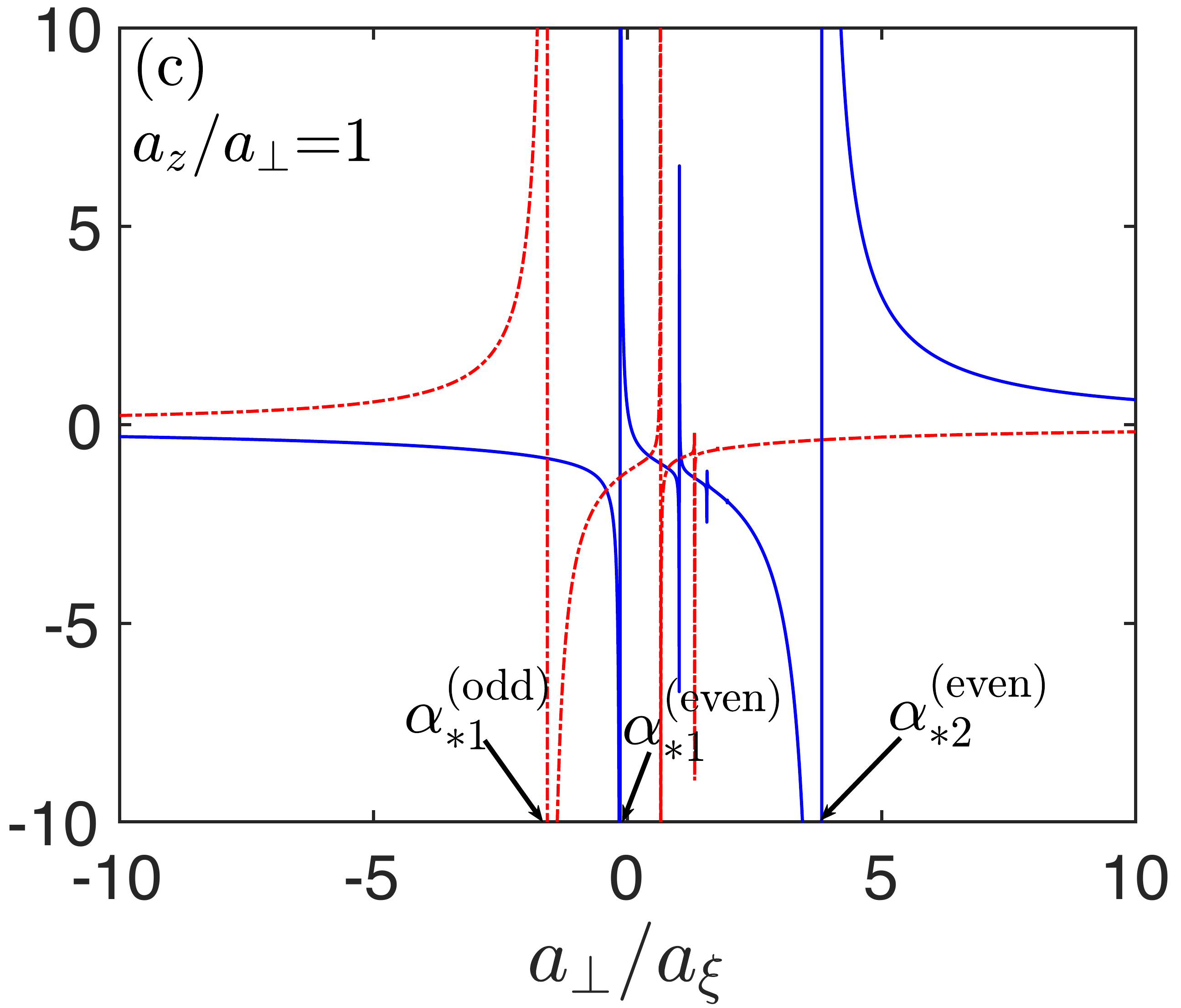}
\includegraphics[width=5.5cm]{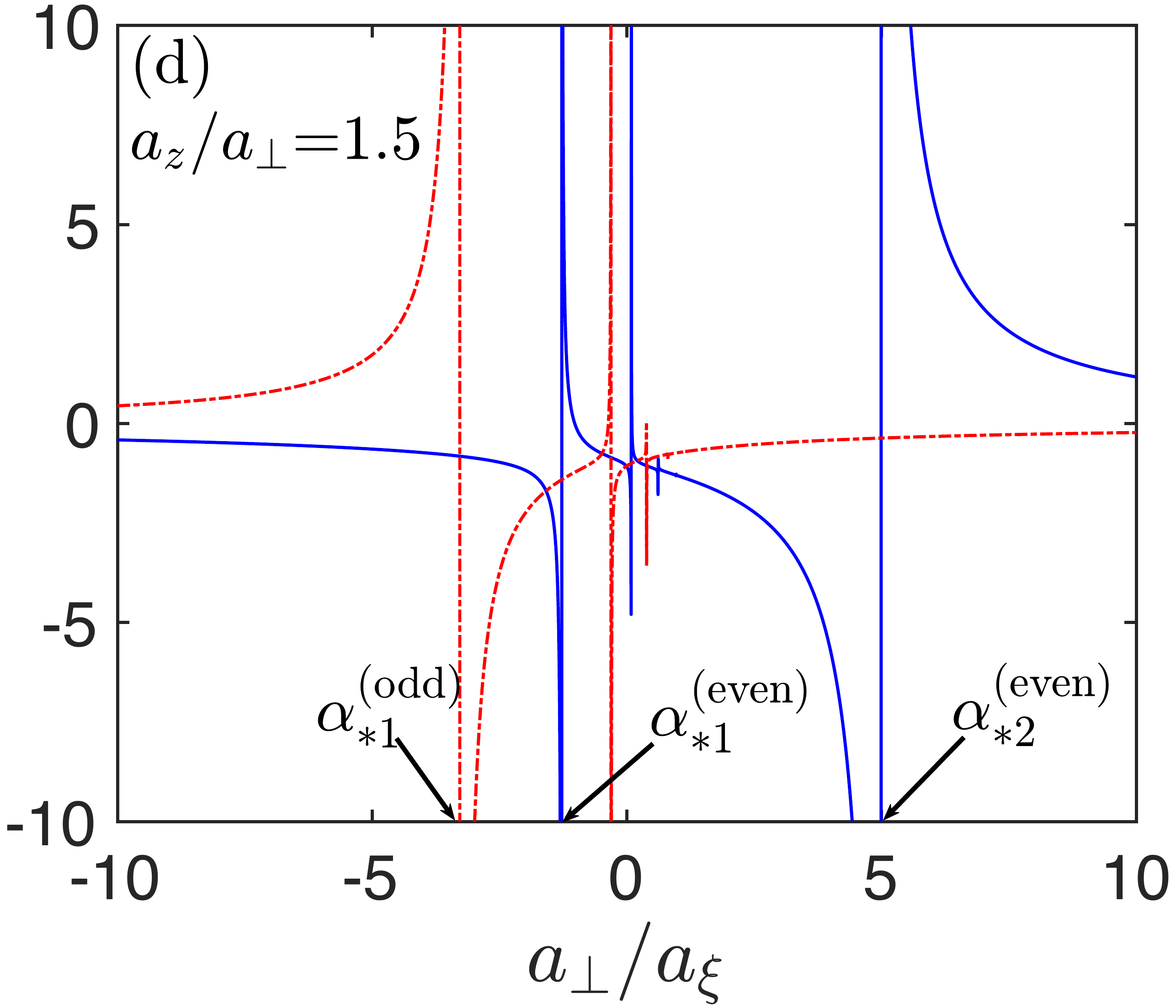}
\includegraphics[width=5.5cm]{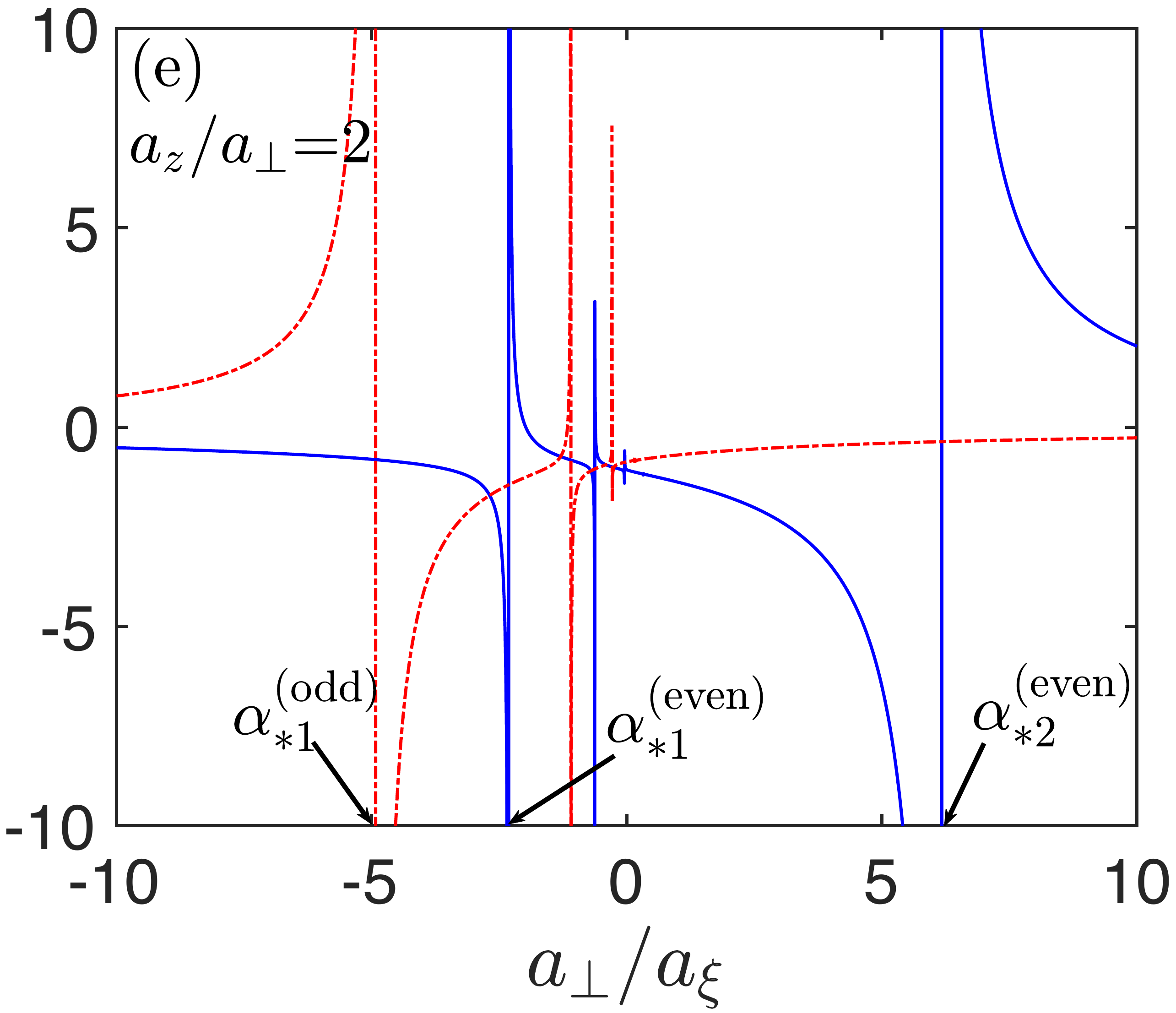}
\includegraphics[width=5.5cm]{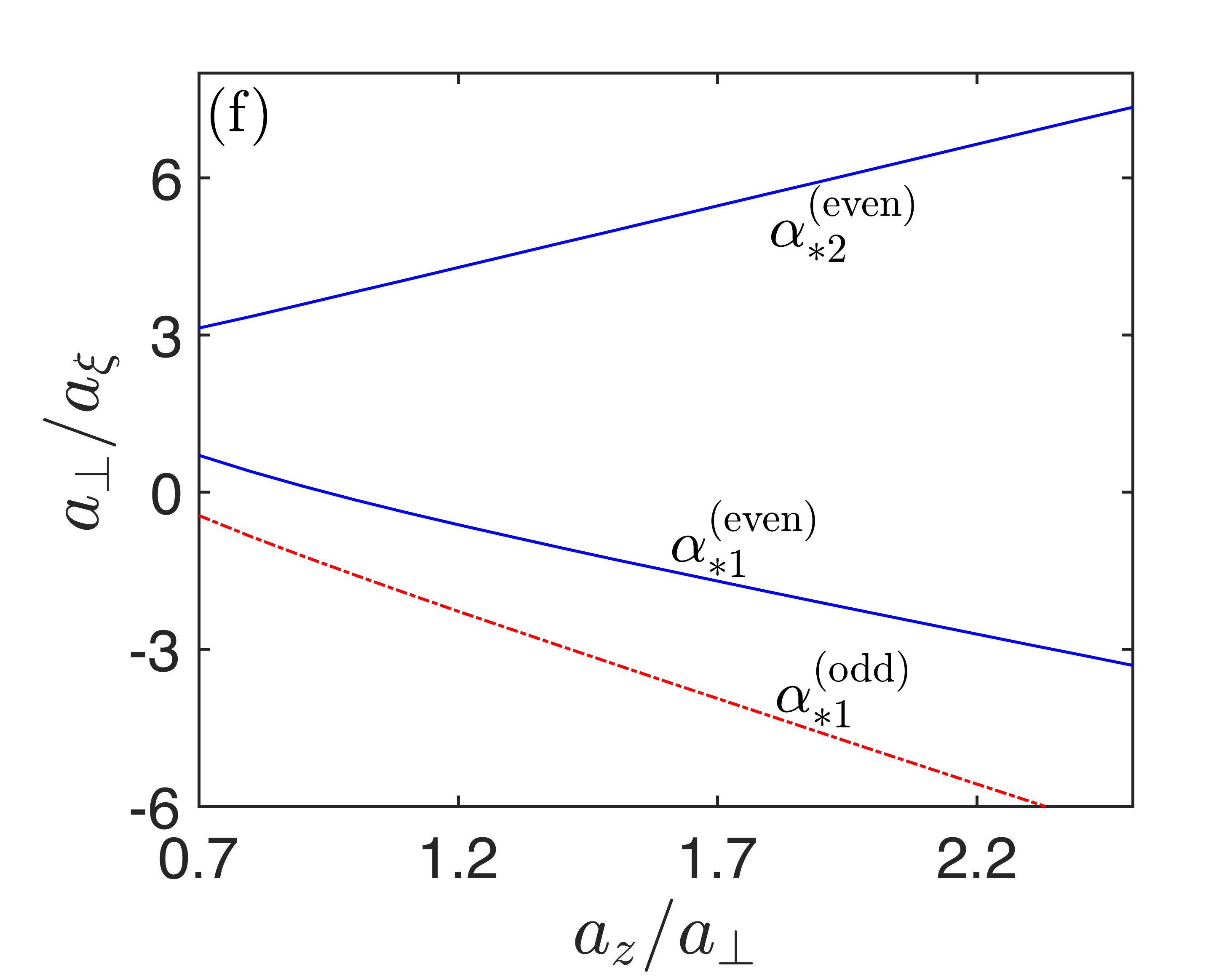}
\caption{(color online) {\bf (a-e):} Effective 1D interaction strength $g_\xi^{\rm (even)}$ (blue solid line) and $g_\xi^{\rm (odd)}$ (red dashed-dotted line) as functions of $a_\perp/a_\xi$, for $a_z/a_\perp=0.5$ (a), $a_z/a_\perp=0.7$ (b),
$a_z/a_\perp=1$ (c), $a_z/a_\perp=1.5$ (d) and $a_z/a_\perp=2$ (e). {\bf (f):} The locations $\alpha^{\rm (even)}_{\ast1}$ and $\alpha^{\rm (even)}_{\ast2}$ of the two broadest even-wave CIR, as well as the location $\alpha^{\rm (odd)}_{\ast1}$ of the broadest odd-wave CIR,
for various $a_z/a_\perp$.}\label{g1d}
\end{center}
\end{figure*}

In our calculation, we first numerically solve Eq. (\ref{eta2}) and
obtain $\eta_{\xi}(z)$, and then substitute our results into Eq.
(\ref{fxi}) and obtain the scattering amplitudes $f_{\xi}^{{\rm even/odd}}(k)$.

\subsubsection{Low-energy behaviors of $f_{\xi}^{{\rm even/odd}}(k)$}

Furthermore, in the low-energy limit $k\rightarrow0$ the behaviors of the scattering amplitudes $f_{\xi}^{{\rm even/odd}}(k)$ are given by \cite{olshanii,fodd1,fodd2,fodd3,fodd4,fodd5,fodd6}
\begin{eqnarray}
f_{\xi}^{{\rm even}}(k)&\approx&-\frac{1}{1+ika_{\xi}^{{\rm (even)}}};\label{fea}\\
\ f_{\xi}^{{\rm odd}}(k)&\approx&\frac{-ik}{ik+\frac{1}{a_{\xi}^{\rm (odd)}}}, \label{sc}
\end{eqnarray}
where 
\begin{eqnarray}
a_{\xi}^{{\rm (even)}}&\equiv&\lim_{k\rightarrow0}\frac{i}{k}\left[1+\frac{1}{f_{\xi}^{{\rm even}}(k)}\right];\label{aeff}\\
a_{\xi}^{{\rm (odd)}}&\equiv&\lim_{k\rightarrow0}\frac{i}{k}\left[1+\frac{1}{f^{\rm odd}(k)}\right]^{-1}\label{ao}
\end{eqnarray}
are the effective 1D scattering lengths for the even and odd waves, respectively.

\subsubsection{Effective 1D interaction}

In addition, the low-energy scattering amplitudes in Eq. (\ref{fea}) and Eq. (\ref{sc}) can be reproduced
by the effective 1D interaction $g_{\xi}^{\rm(even)}\delta(z_g)(z_g){\hat d}_{\rm e}+g_{\xi}^{\rm(odd)}\delta^\prime(z_g){\hat d}_{\rm o}$,
with ${\hat d}_{\rm e,o}$ being defined in Eqs. (\ref{de}, \ref{do}), and the intensities $g_{\xi}^{{\rm (even/odd)}}$ being given by \cite{olshaniiodd}
\begin{eqnarray}
g_{\xi}^{{\rm (even)}}&=&-\frac{1}{a_{\xi}^{{\rm (even)}}},\label{geff2a}\\ 
g_{\xi}^{{\rm (odd)}}&=&-{a_{\xi}^{{\rm (odd)}}}.\label{geff2}
\end{eqnarray}
Therefore, when the scattering amplitude $f_{\xi}^{{\rm even/odd}}(k)$ are obtained,
we can calculate the effective 1D scattering lengths $a_{\xi}^{{\rm (even/odd)}}$ via Eq. (\ref{aeff}) and Eq. (\ref{ao}),
and then derive the effective 1D interaction intensity $g_{\xi}^{{\rm (even/odd)}}$
via Eq.  (\ref{geff2a}) and Eq. (\ref{geff2}).

\section{Results and Analysis}

In the above section we show our approach for the numerical calculation for the strengths $g_{\pm}^{\rm(even/odd)}$  of the effective 1D interaction $V_{\pm}^{({\rm eff})}$. 
In this section, we illustrate our results and study the CIRs for our system, and then compare our results with the experimental observations in Ref. \cite{blochexp}.

\subsection{Locations and Widths of the CIRs}

As shown above,
$g_{+}^{\rm(even/odd)}$ and $g_{-}^{\rm(even/odd)}$ are given by the scattering amplitudes for the same scattering problem with different 3D scattering lengths $a_+$ and $a_-$, respectively. As a result, an immediate dimensional analysis yields that 
\begin{eqnarray}
g_{\xi}^{(l)}={a_\xi}^{\lambda_l}S_{l}\left(\frac{a_\perp}{a_\xi},\frac{a_z}{a_\perp}\right),\label{ffun}
 \end{eqnarray}
for $l=$(even, odd) and $\xi=(+,-)$, with $\lambda_{\rm even}=-1$, $\lambda_{\rm odd}=+1$, and  $S_{\rm even}(\alpha,\beta)$ and $S_{\rm odd}(\alpha,\beta)$ being $\xi$-independent universal functions which can be obtained via numerical calculation shown in Sec. II. This result shows that for a system with fixed 3D scattering length $a_\xi$, the control effect of the parameters $a_z$ and $a_\perp$ for $g_{\xi}^{\rm(even/odd)}$ can be described by the two dimensionless parameters $a_\perp/a_\xi$ and $a_z/a_\perp$ with the following clear physical meanings. The absolute value of $a_\perp/a_\xi$ describes the intensity of the transverse confinement. Explicitly, $|a_\perp/a_\xi|$ is larger for weaker transverse confinement. Similarly, 
the ratio $a_z/a_\perp$ describes the relative intensity of the axial and transverse confinement. 

In addition, Eq. (\ref{ffun}) shows the condition for the CIR in the even and odd wave channels
can be expressed as
\begin{eqnarray}
\left(\frac{a_z}{a_\xi},\frac{a_z}{a_\perp}\right)=\left(\alpha^{\rm (even)}_\ast,\beta^{\rm (even)}_\ast\right),\label{cc1}
\end{eqnarray}
and
\begin{eqnarray}
\left(\frac{a_z}{a_\xi},\frac{a_z}{a_\perp}\right)=\left(\alpha^{\rm (odd)}_\ast,\beta^{\rm (odd)}_\ast\right),
\label{cc2}
\end{eqnarray}
respectively, with $(\alpha^{\rm (even/odd)}_\ast,\beta^{\rm (even/odd)}_\ast)$ being any singularity of the function $S_{\rm even/odd}(\alpha,\beta)$. It is clear that the values of $(\alpha^{\rm (even/odd)}_\ast,\beta^{\rm (even/odd)}_\ast)$ are $\xi$-independent.
When the condition in Eq. (\ref{cc1}) or Eq. (\ref{cc2}) is satisfied for a specific $\xi$, we have $g_\xi^{\rm (even)}=\infty$ or $g_\xi^{\rm (odd)}=\infty$. According to Eqs. (\ref{veff2}, \ref{omed}, \ref{ome}), in this case the strength of the effective 1D spin-exchange interaction, i.e., $\Omega^{\rm (even)}$ or $\Omega^{\rm (odd)}$, also diverges.

Now we investigate the locations and widths of the CIRs. To this end, in Fig.~\ref{g1d} (a-e) we illustrate the dependence of $g_{\xi}^{\rm(even)}$ (in units of $1/a_z$) and $g_{\xi}^{\rm(odd)}$ (in units of $a_z$) on ${a_\perp}/{a_\xi}$, for given values of  ${a_z}/{a_\perp}$. In addition, in Fig.~\ref{g1d} (f) we plot the locations $\alpha^{\rm (even)}_{\ast 1}$ and $\alpha^{\rm (even)}_{\ast 2}$ of the two broadest even-wave CIRs, as well as the location $\alpha^{\rm (odd)}_{\ast 1}$ of the broadest odd-wave CIR, as functions of ${a_z}/{a_\perp}$. The results in these figures can be summarized and understood as follows:

(A) Multiple CIRs can appear for both even- and odd-wave channels. This result is qualitatively consistent with our previous work with the pure-1D model \cite{our2}. As stated in \cite{our2}, it can be explained as the result of the coupling between the center-of-mass motion and the relative motion of the two atoms in the $z$-direction. Similar multi-resonance phenomena were also found in other  scattering problems between two ultracold atoms, where the center-of-mass motion is coupled to the relative motion \cite{castin,tan-mixd-d,rccouple1,rccouple2,rccouple3,rccouple4,rccouple5,rccouple6,rccouple7,tannishida}.

(B) For the odd partial wave, the broadest CIR is the one located at the lower end of $a_\perp/a_\xi$. The location is denoted as $a_\perp/a_\xi=\alpha^{\rm (odd)}_{\ast1}$. Other odd-wave CIRs become more and more narrow when $a_\perp/a_\xi$ increases. In addition, as shown in Fig.~\ref{g1d} (c-e), for ${a_z}/{a_\perp}\gtrsim 0.7$, we have $\alpha^{\rm (odd)}_{\ast1}<0$, i.e., the broadest odd-wave CIR can appear only when the $s$-wave scattering length $a_\xi$ is negative.

(C) For the even partial wave, when ${a_z}/{a_\perp}$ is small (e.g., ${a_z}/{a_\perp}=0.5$, as shown in Fig.~\ref{g1d} (a)), the situation is similar as the odd partial wave, i.e., the broadest CIR is the one located at the lower end of $a_\perp/a_\xi$. When the value of ${a_z}/{a_\perp}$ becomes large (Fig.~\ref{g1d} (b)), some narrow CIRs, which appear for relatively large $a_\perp/a_\xi$, gradually merge with each other and form another broad CIR. Furthermore, in the parameter region with ${a_z}/{a_\perp}\gtrsim 1$ (Fig.~\ref{g1d} (c-e)), there are always two relatively broad CIRs which are located as  $a_\perp/a_\xi=\alpha^{\rm (even)}_{\ast1}$ and $\alpha^{\rm (even)}_{\ast2}$, with $\alpha^{\rm (even)}_{\ast1}<0$ and $\alpha^{\rm (even)}_{\ast2}>0$. In addition, many relatively narrow CIRs can occur  for $\alpha^{\rm (even)}_{\ast1}<a_\perp/a_\xi<\alpha^{\rm (even)}_{\ast2}$. Thus, when ${a_z}/{a_\perp}\gtrsim 1$ a broad CIR,  which is usually very advantageous for the control of inter-atomic interaction, can always be realized for the systems with either positive or negative 3D scattering length $a_\xi$.

(D) As shown in Fig.~\ref{g1d} (f), for ${a_z}/{a_\perp}\gtrsim 0.7$ the absolute values of $\alpha^{\rm odd}_{\ast 1}$ and $\alpha^{\rm even}_{\ast 1,2}$, i.e., the locations of the broadest odd-wave CIR and the two broadest even-wave CIR,  almost linearly increase with ${a_z}/{a_\perp}$. Thus, for either positive or negative $a_\xi$, by increasing the ratio ${a_z}/{a_\perp}$ one can always realize a broad even-wave CIR via the confinements with larger characteristic lengths of $a_\perp$ and $a_z$. 
These confinements can be created via weaker laser beams, and thus are more feasible to be prepared in the experiments. 
Similarly, when $a_\xi$ is negative one can also realize a broad odd-wave CIR in these confinements.
Nevertheless, in realistic systems ${a_z}/{a_\perp}$ cannot be infinitely increased.  That is because, when  ${a_z}/{a_\perp}\rightarrow\infty$, we have either ${a_z}\rightarrow\infty$ or ${a_\perp}\rightarrow 0$. In the former case,  we also have $\omega_z\rightarrow0$  and thus
the low-temperature condition $T\ll \omega_z/k_B$ would be violated. In the latter case, the transverse confinement has to be realized via very strong laser beams, which is formidable in experiments.





\subsection{Theory-Experiment Comparison }

Now we compare our theoretical results with the recent experiment of ultracold  $^{173}$Yb atoms \cite{blochexp}. In this experiment the transverse confinement for both the two atoms and the axial confinement for the $e$-atom are realized via a 2D optical lattice with magic wave lengths $\lambda_\perp=759$nm and a 1D optical lattice with wave length $\lambda_z=680$nm, respectively. The explicit potentials of these two lattices are given by
\begin{eqnarray}
V_{\rm lattice}^{(\perp)}&=&U_{\perp}\sum_{j=e,g}\left[\cos^2\left(2\pi x_j/\lambda_\perp\right)+\cos^2\left(2\pi y_j/\lambda_\perp\right)\right];\nonumber\\
\label{vlattice}\\
V_{\rm lattice}^{(z)}&=&U_{z}\cos^2(2\pi z_e/\lambda_z),\label{vz}
\end{eqnarray}
where $x_j$ and  $y_j$ ($j=e,g$) are the $x$- and $y$- coordinates of the $j$-atom, respectively, and
$U_{z}$ and $U_\perp$ are the intensities of the lattices. By expanding these two potentials around the minimum points we can obtain the harmonic trapping potentials shown in Eqs. (\ref{hperp}) and (\ref{he}). The characteristic lengths $a_z$ and $a_\perp$ are given by
\begin{eqnarray}
a_\perp=\frac{\sqrt{\lambda_\perp}}{2^{\frac14}\sqrt{\pi} U_\perp^{\frac14}};\
a_z=\frac{\sqrt{\lambda_z}}{2^{\frac34}\sqrt{\pi} U_z^{\frac14}},
\end{eqnarray}
and can be tuned via the intensities $U_{z}$ and  $U_{\perp}$. 
In the experiments of Ref.~\cite{blochexp}, the $e$-atoms and $g$-atoms are initially prepared in the states $|\downarrow\rangle_e$ and $|\uparrow\rangle_g$, respectively. After a finite holding time, the spin of some $e$-atoms are flipped to state $|\uparrow\rangle_e$ by the effective spin-exchange interaction. The number $N_{e\uparrow}$ of the spin-flipped $e$-atoms is measured. This number is supposed to be positively correlated with the absolute value of the effective 1D spin-exchange  intensity $\Omega^{\rm (even/odd)}$, which is given by $\Omega^{\rm (even/odd)}=(g_-^{\rm(even/odd)}-g_+^{\rm(even/odd)})/2$ in our theory, as shown in Eqs. (\ref{veff2}, \ref{ome}). Explicitly, when the system is around a CIR of either $g_+^{\rm(even/odd)}$ or $g_-^{\rm(even/odd)}$,  
the atom number $N_{e\uparrow}$ should be resonantly enhanced.

\begin{figure}[htp]
\begin{center}
\includegraphics[width=6.5cm]{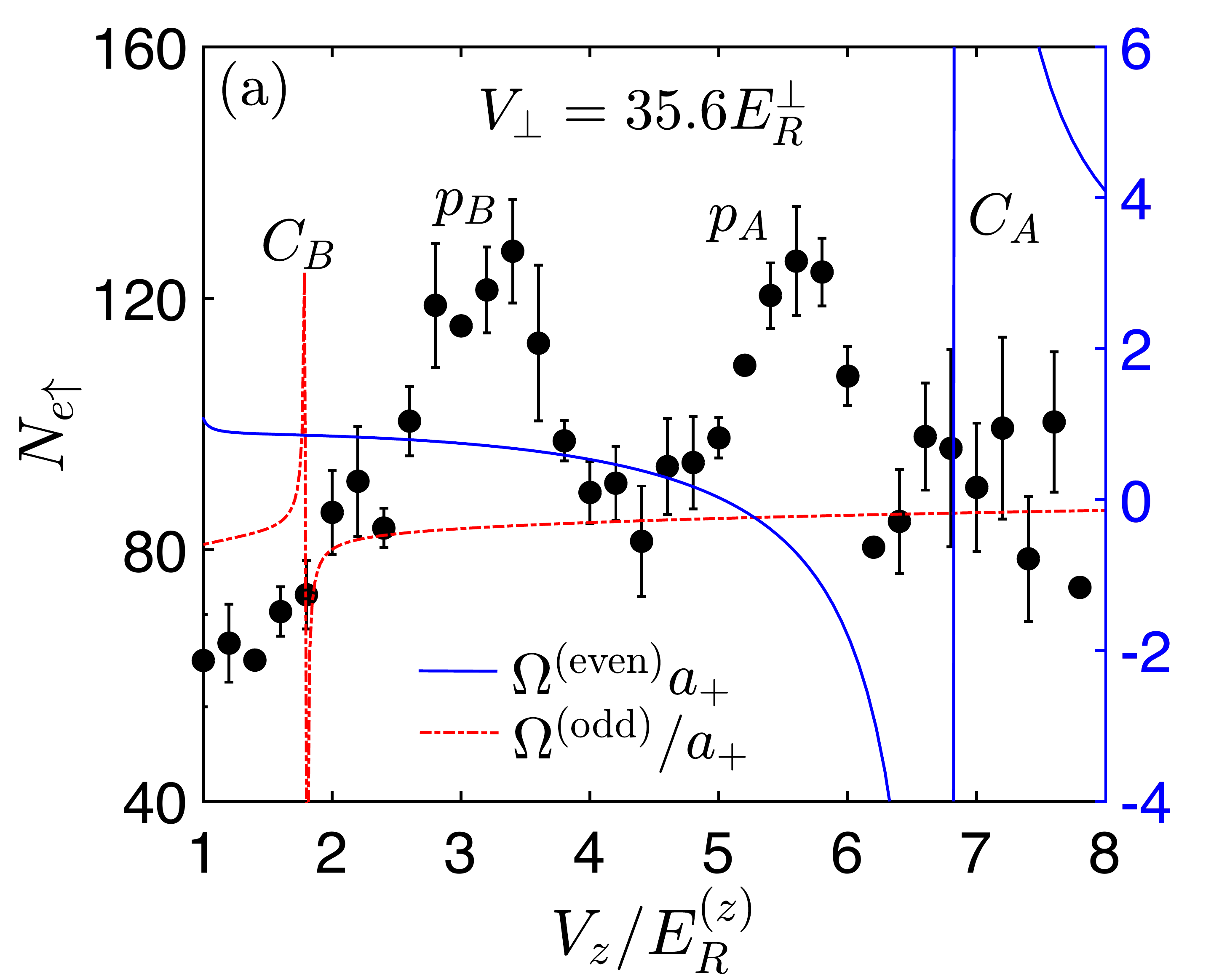}
\includegraphics[width=6.5cm]{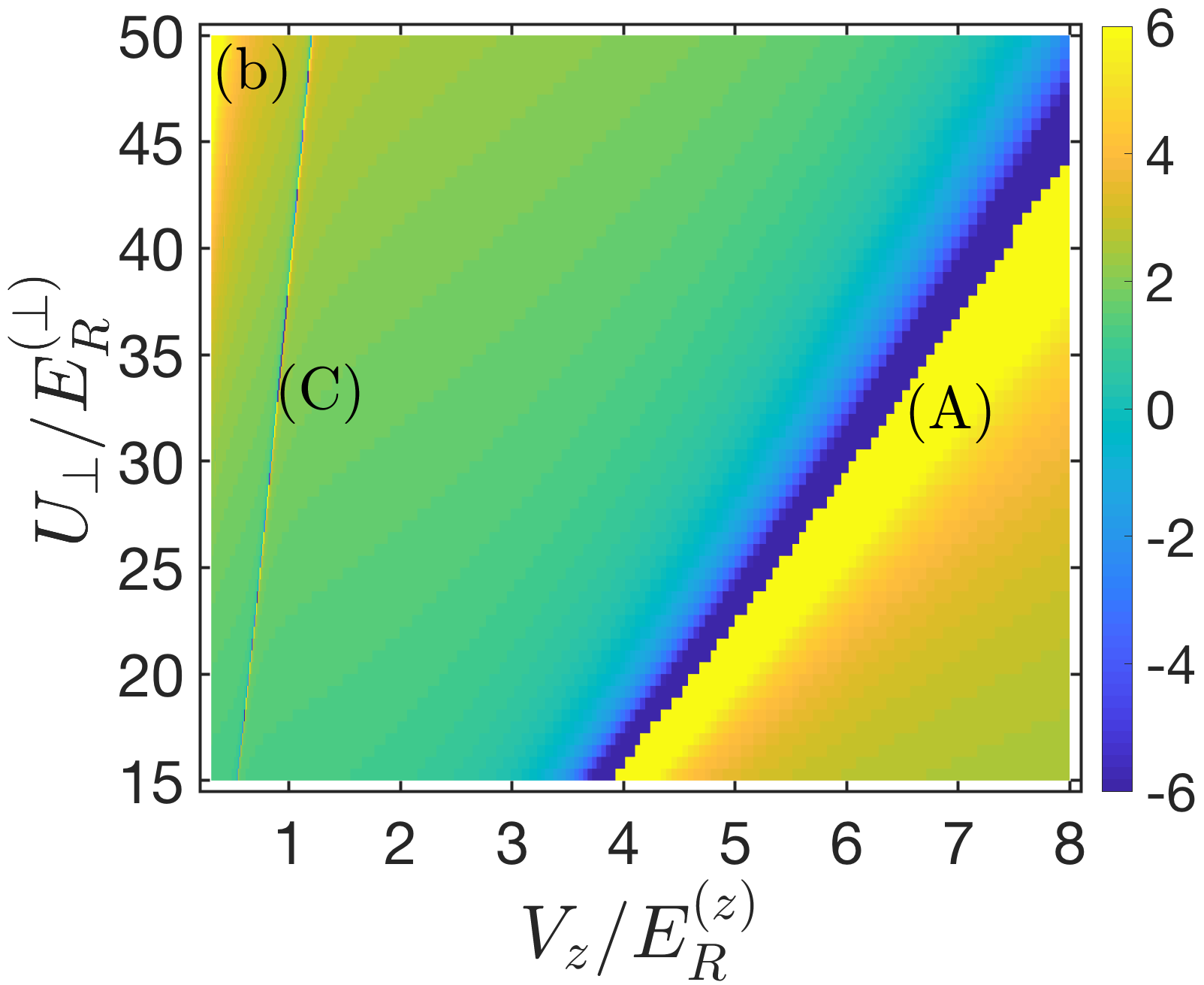}
\includegraphics[width=6.5cm]{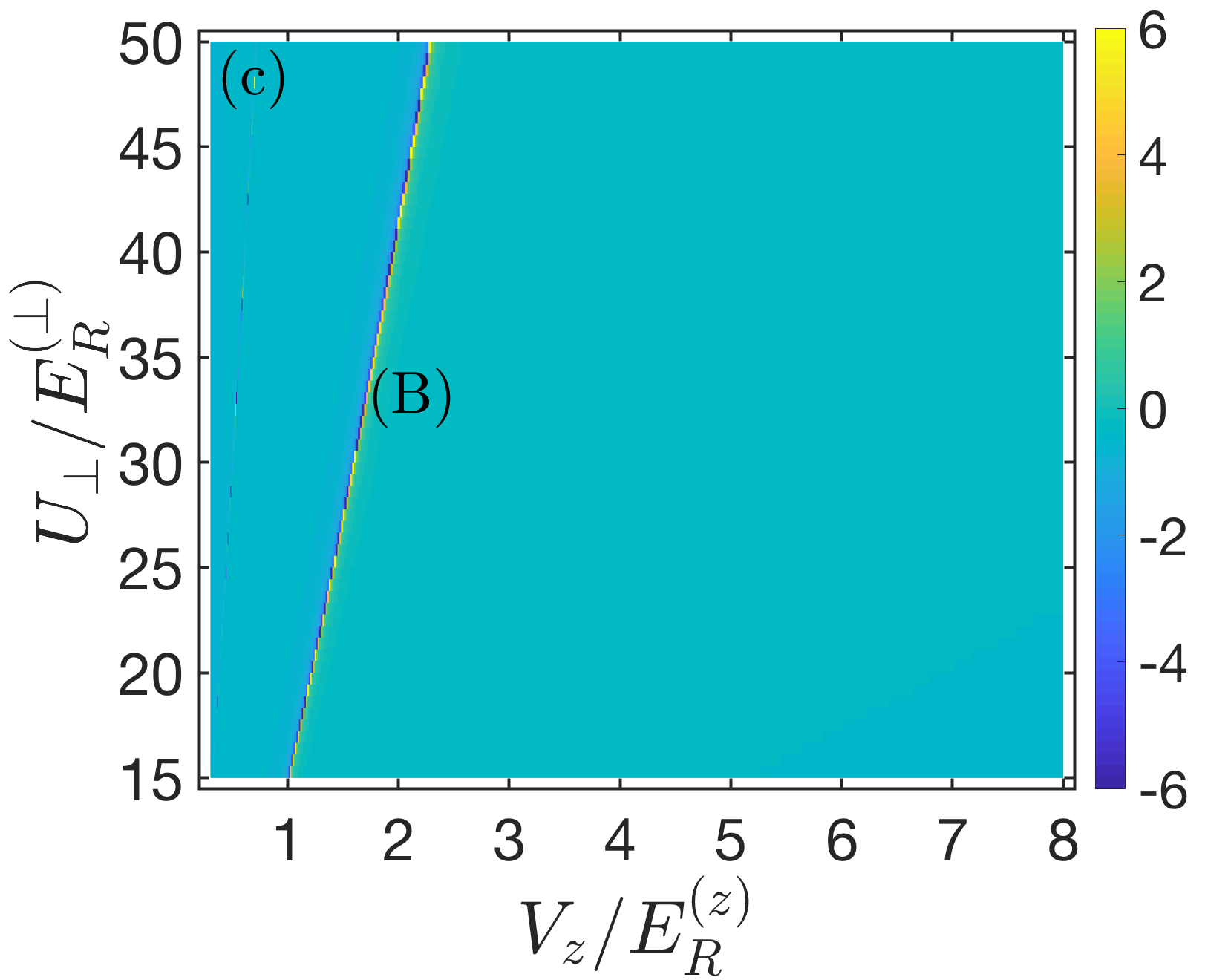}
\caption{(color online) Comparison between our theory with the  experimental results of Ref. \cite{blochexp}. In our calculations we take $a_+=1878a_0$ and $a_-=216a_0$ \cite{ofrexp2}. 
{\bf (a)}: Black dots with error bar: the number $N_{e\uparrow}$ of spin-flipped $e$-atoms  observed in the experiment, for the cases with $U_\perp=35.6E_R^{(\perp)}$. The datas are taken from Fig. 3(b) of Ref. \cite{blochexp}. Blue solid line: the 
effective 1D even-wave spin-exchange  intensity $\Omega^{\rm (even)}$ (in units of $1/a_+$) given by our calculation. Red dashed-dotted line: the 
effective 1D odd-wave spin-exchange  intensity $\Omega^{\rm (odd)}$ (in units of $a_+$) given by our theoretical calculation. 
As shown in our main text, the horizontal ordinate $V_z$ is defined as $V_z\equiv U_z/3.3$, with $U_z$ being defined in Eq. ~(\ref{vz}).
{\bf (b)}: Theoretically calculated  $\Omega^{\rm (even)}$ (in units of $1/a_+$), as a function of $V_z$ and $U_\perp$.
{\bf (c)}: Theoretically calculated  $\Omega^{\rm (odd)}$ (in units of $a_+$), as a function of $V_z$ and $U_\perp$.
}
 \label{exp}
\end{center}
\end{figure}

In Fig.~\ref{exp}(a) we compare the  experimentally observed atom number $N_{e\uparrow}$ and the effective spin-exchange interaction intensity $\Omega^{\rm (even)}$ and $\Omega^{\rm (odd)}$ obtained from our theory, for the cases with $U_\perp=35.6E_{\rm R}^{(z)}$, where $E_{\rm R}^{(z)}=2\pi^2/\lambda_{z}^2$ and  $E_{\rm R}^{(\perp)}=2\pi^2/\lambda_{\perp}^2$ are the recoil energies of the lattices. To be consistent with Ref.~\cite{blochexp}, in the figure we chose the horizontal ordinate to be $V_z\equiv U_z/3.3$. 
As shown in this figure, a even-wave CIR  ($C_{A}$) and  an odd-wave CIR  ($C_{B}$) 
are found by our  calculations, with the positions being close to the experimentally observed peaks $p_{A}$ and $p_{B}$ of $N_{e\uparrow}$, respectively. According to this result, the  peaks $p_{A}$ and $p_{ B}$ are due to  CIRs in different partial-wave channels.
In addition, there are some  difference between the position of CIR $C_{A,B}$ and the observed peak $p_{A,B}$. The difference between the positions of $C_{A}$ and $p_{A}$ may due to the fact that in our calculation the shallow lattice experienced by the $g$-atom is ignored. On the other hand, the difference between the positions of $C_{B}$ and $p_{B}$ may due to the fact that, in the region of $C_{B}$ and $p_{B}$ ($V_z\lesssim 3E_{\rm R}^{(z)}$, i.e., $U_z\lesssim 9.9E_{\rm R}^{(z)}$) the depth of the trapping potential $V_{\rm lattice}^{(z)}$ experienced by the $e$-atom is so weak that the harmonic approximation for this potential does not work very well.

In Fig.~\ref{exp}(b) and \ref{exp}(c) we further illustrate  $\Omega^{\rm (even)}$ and $\Omega^{\rm (odd)}$ given by our calculation, respectively, for the parameter region 
$U_\perp/E_{\rm R}^{(\perp)}\in[15,50]$ and $V_z/E_{\rm R}^{(z)}\in[0.3,8]$ (i.e., $U_z/E_{\rm R}^{(\perp)}\in[1,26.4]$). 
In these figures, the locations of the CIRs are the places where the color suddenly changes from deep blue to yellow, where $|\Omega^{\rm (even/odd)}|$ is very large and the sign of $\Omega^{\rm (even/odd)}$ suddenly changes. 
Two even-wave CIR branches (A), (C) and one odd-wave CIR branch (B) are illustrated. Our calculations show that all of them are caused by the CIRs of $g_+^{\rm(even/odd)}$. 
 The experimental results for the parameter region of Fig.~\ref{exp}(b, c) 
are given in Fig.~3 of Ref. \cite{blochexp}, where the lattice depth $U_\perp$ is denoted as $V_\perp$. It is shown that the even-wave CIR branch (A) and odd-wave CIR branch (B) are clearly observed at the locations which are close to our theoretical results. In addition, similar as in Fig.~\ref{exp}(a),
The quantitative shifts between the theoretical prediction to the experimental observation for 
these two CIR branches are due to the ignorance of the shallow lattice for the $g$-atom, as well as
the weakness of the trapping lattice potential $V_{\rm lattice}^{(z)}$ for the $e$-atom in the region of CIR (B). Furthermore, the even-wave CIR branch (C) is not experimentally detected. That is very possibly because this branch is too narrow, as shown in Fig.~\ref{exp}(b).



\section{Summary and Outlook}
We exactly solved the scattering problem between two alkali-earth (like) atoms, i.e., one $g$-atom freely moving in a quasi-1D tube and another  $e$-atom localized by a 3D harmonic trap. Our solutions show that with the help of the even- or odd-wave CIRs in this system, the effective 1D spin-exchange interaction can be resonantly controlled via the characteristic lengths $a_z$ and $a_\perp$ of the confinements.  When $a_z/a_\perp$ is larger, the relatively broad CIRs can be realized in weaker confinements. Our results reveal that the two  CIR branches which are observed in 
in the recent experiment in Ref. \cite{blochexp} are due to an  even-wave and an odd-wave CIRs, respectively. To our knowledge, in previous studies for the Kondo effect, most of the attentions were paid to the systems with only an even-wave 1D interaction. The system with the resonant odd-wave spin-exchange interaction were not studied so much. Our results show that these systems can be experimentally realized in the ultracold gases of alkali-earth (like) atoms.

As shown above, in the experiments a shallow axial lattice potential for the $g$-atom can also be induced by a laser beam which is used to confine the axial motion of the  $e$-atom.
In future, we will further study the effect induced by this shallow lattice potential.

\begin{acknowledgments}
This work is 
is supported in part by the National Key R$\&$D Program of China (Grant No. 2018YFA0307601 (RZ), 2018YFA0306502 (PZ)), NSFC Grant No. 11804268 (RZ), 11434011(PZ), No. 11674393
(PZ), as well as the Research Funds of Renmin University of China
under Grant No. 16XNLQ03(PZ).
\end{acknowledgments}


\appendix

\begin{widetext}
\section{Partial-Wave Analysis}

In this appendix we prove that  $f_{\xi}^{{\rm even/odd}}(k)$ given by Eq. (\ref{fxi}) are just the scattering amplitudes for the even/odd partial waves.

 As shown in Sec. II. B, our system in invariable under the total  reflection operation 
$
T:\{z_g\rightarrow-z_g,z_e\rightarrow-z_e\}.
$
Thus, the subspaces ${\cal H}_{\rm even}\equiv\{\Psi|T\Psi=+\Psi\}$ (corresponding to the parity ${\mathbb P}=+1$, or even partial wave) and ${\cal H}_{\rm odd}\equiv\{\Psi|T\Psi=-\Psi\}$ (corresponding to ${\mathbb P}=-1$, or odd partial wave) are
 invariant subspaces of the complete Hamiltonian $H$. Furthermore, the incident state $\Psi^{(0)}(\bm{\rho},z_{e},z_{g})$ defined in Eq. (\ref{psik0}) can be expressed as
 \begin{eqnarray}
 \Psi^{(0)}=  \Psi_{\rm even}^{(0)}+ \Psi_{\rm odd}^{(0)},
 \end{eqnarray}
 where $ \Psi_{\rm even/odd}^{(0)}\in {\cal H}_{\rm even/odd}$ and are defined as
  \begin{eqnarray}
 \Psi_{\rm even}^{(0)}(\bm{\rho},z_{e},z_{g})&=&  \frac{\cos(kz_{g})}{\sqrt{2\pi}}\chi_{n_{\perp}=0,m_{z}=0}(\bm{\rho})\phi_{n_{z}=0}(z_{e});\\
  \Psi_{\rm odd}^{(0)}(\bm{\rho},z_{e},z_{g})&=& i \frac{\sin(kz_{g})}{\sqrt{2\pi}}\chi_{n_{\perp}=0,m_{z}=0}(\bm{\rho})\phi_{n_{z}=0}(z_{e}).
 \end{eqnarray}
 Therefore, the scattering wave function $\Psi_{\xi}(\bm{\rho},z_{e},z_{g})$ corresponding to the incident state $\Psi^{(0)}$, which we studied in Sec. II. B, can be expressed as
  \begin{eqnarray}
 \Psi_\xi(\bm{\rho},z_{e},z_{g})=  \Psi_\xi^{\rm even}(\bm{\rho},z_{e},z_{g})+ \Psi_\xi^{\rm odd}(\bm{\rho},z_{e},z_{g}),\label{a5}
 \end{eqnarray}
 with $\Psi_\xi^{\rm even/odd}\in{\cal H}_{\rm even/odd}$ being the scattering wave functions  corresponding to the incident states $ \Psi_{\rm even/odd}^{(0)}$, i.e., the scattering states in the even/odd partial-wave channel. In addition, using the analysis which are similar as Sec. II. B, we can find that 
 \begin{eqnarray}
\Psi_\xi^{\rm even/odd}(\bm{\rho},z_{e},|z_{g}|\rightarrow\infty)= \Psi_{\rm even/odd}^{(0)}(\bm{\rho},z_{e},z_{g})
+f_{\xi}^{{\rm even/odd}}(k)\Lambda_{\rm even/odd}e^{ik|z_{g}|}\chi_{n_{\perp}=0,m_{z}=0}(\bm{\rho})\phi_{n_{z}=0}(z_{e}),\label{feeoo}
\end{eqnarray}
where $\Lambda_{\rm even}=1$ and $\Lambda_{\rm odd}={\rm sign}(z_{g})$, and $f_{\xi}^{{\rm even/odd}}(k)$ being given by Eq. (\ref{fxi}). 
In the derivation of this result, we have used $\eta_\xi(z')=\eta^{\rm even}_\xi(z')+\eta^{\rm odd}_\xi(z')$, with $\eta_\xi(z')$ being defined in Eq. (\ref{etaz}) and $\eta^{\rm even/odd}_{\xi}(z')=\left.\frac{\partial}{\partial z_{r}}\left[z_{r}\Psi_{\xi}^{\rm even/odd}\left({\bf 0},z'-\frac{z_{r}}{2},z'+\frac{z_{r}}{2}\right)\right]\right|_{z_{r}\rightarrow0^{+}}$, and the facts that $\eta^{\rm even/odd}_\xi(z')$ are even/odd functions of $z'$.

Eq. (\ref{feeoo}) shows that $f_{\xi}^{{\rm even/odd}}(k)$ are nothing but the scattering amplitudes for the even/odd partial wave.

\section{Integral Equation for $\eta_{\xi}(z)$}

\label{STM} 
In this appendix we derive the integral equation for the regularized wave function
$\eta_{\xi}(z)$ defined in Eq. (\ref{etaz}). 

\subsection{The Green's function $G_{E}({\bf 0},z_{e},z_{g};{\bf 0},z^{\prime},z')$}

The Green's function $G_{E}({\bf 0},z_{e},z_{g};{\bf 0},z^{\prime},z')$ is very important for our calculation.
Therefore, here we first re-express this function into convenient forms. Using Eq. (\ref{ge2})  and the fact $|\chi_{n_{\perp},m_{z}=0}({\bf 0})|^{2}=\omega_{\perp}/(2\pi)$,
we have
\begin{equation}
G_{E}({\bf 0},z_{e},z_{g};{\bf 0},z^{\prime},z')=\frac{\omega_{\perp}}{2\pi}\sum_{n_{\perp}=0,2,4,6,...}g\left[E-(n_{\perp}+1)\omega_{\perp};z_{e},z_{g},z',z'\right],\label{a1}
\end{equation}
where the function $g({\cal E};z_{e},z_{g};z_{e}',z_{g}')$ is the
Green's function for the axial motion of the $g$-atom and the $e$-atom
(i.e., the matrix element of $[{\cal E}+i0^{+}-(H_{0}-H_{\perp})]^{-1}$),
and can be expressed as
\begin{equation}
g({\cal E};z_{e},z_{g};z',z')=\sum_{n_{z}=0,1,2,3,...}\frac{e^{\sqrt{2\left[(n_{z}+\frac{1}{2})\omega_{z}-{\cal E}\right]}|z_{g}-z^{\prime}|}}{\sqrt{2\left[(n_{z}+\frac{1}{2})\omega_{z}-{\cal E}\right]}}\phi_{n_{z}}(z_{e})\phi_{n_{z}}^{*}(z^{\prime}),\label{a2}
\end{equation}
where for the function $\sqrt{q}$ is defined as $\sqrt{q}=i|q|^{\frac{1}{2}}$ for $q<0$.
Eq. (\ref{a1})  yields that
\begin{equation}
G_{E}({\bf 0},z_{e},z_{g};{\bf 0},z^{\prime},z')=\frac{\omega_{\perp}}{2\pi}g\left[E-\omega_{\perp};z_{e},z_{g};z',z'\right]+G_{E^{\prime}}({\bf 0},z_{e},z_{g};{\bf 0},z^{\prime},z'),\label{gre}
\end{equation}
with the energy $E^{\prime}$ being defined as
\begin{equation}
E^{\prime}=E-\omega_{\perp}.\label{ep}
\end{equation}
With the help of Eq. (\ref{gre}) we convert the calculation of $G_{E}$
to the calculations of the functions  $g\left[E-\omega_{\perp};z_{e},z_{g};z',z'\right]$
and $G_{E^{\prime}}$. The former function can be easily calculated
numerically. Furthermore, due to the low-energy assumption shown in
Eq. (\ref{ee}) and Eq. (\ref{enecon}), we know that $E^{\prime}$
is lower than the threshold of $H_{0}$, i.e., $E^{\prime}<\omega_{\perp}+\omega_{z}/2$.
Due to this fact, the integration $-\int_{0}^{\infty}d\beta e^{\beta E^{\prime}}e^{-\beta H_{0}}$
converges, and we have $(E^{\prime}-H_{0})^{-1}=-\int_{0}^{\infty}d\beta e^{\beta E^{\prime}}e^{-\beta H_{0}}$.
Thus, the Green's function $G_{E^{\prime}}$ can be re-expressed as
\begin{equation}
G_{E^{\prime}}({\bf 0},z_{e},z_{g};{\bf 0},z^{\prime},z')=-\int_{0}^{\infty}d\beta e^{\beta E^{\prime}}K_{\beta}({\bf 0},z_{e},z_{g};{\bf 0},z^{\prime},z'),\label{gg}
\end{equation}
with the function $K_{\beta}(\bm{\rho},z_{e},z_{g};\bm{\rho^{\prime}},z_{e}^{\prime},z_{g}')$
being the imaginary-time propagator of the free Hamiltonian $H_{0}$,
and can be expressed as
\begin{eqnarray}
K_{\beta}(\bm{\rho},z_{e},z_{g};\bm{\rho^{\prime}},z_{e}^{\prime},z_{g}') & = & \langle\bm{\rho},z_{e},z_{g}|e^{-\beta H_{0}}|\bm{\rho^{\prime}},z_{e}^{\prime},z_{g}^{\prime}\rangle\nonumber \\
 & = & K_{\beta}^{(\perp)}(\bm{\rho},\bm{\rho^{\prime}})K_{\beta}^{(e)}(z_{e},z_{e}^{\prime})K_{\beta}^{(g)}(z_{g},z_{g}^{\prime}),\label{kk}
\end{eqnarray}
where $K_{\beta}^{(\perp)}$, $K_{\beta}^{(e)}$ and $K_{\beta}^{(g)}$
are the propogators of the two-atom transverse relative motion (two-dimensional
harmonic oscillator with mass $1/2$ and frequency $\omega_{\perp}$),
the axial motion of the $e$-atom (1D harmonic oscillator with mass
$1$ and frequency $\omega_{z}$) and the axial motion of the $g$-atom
(1D free particle with mass $1$), respectively. Explicitly, we have
\begin{eqnarray}
K_{\beta}^{(\perp)}(\bm{\rho},\bm{\rho^{\prime}}) & = & \left[\frac{\omega_{\perp}}{4\pi\sinh(\omega_{\perp}\beta)}\right]\exp\left\{ -\frac{\omega_{\perp}\left[\left(|\bm{\rho}|^{2}+|\bm{\rho^{\prime}}|^{2}\right)\cosh(\omega_{\perp}\beta)-2\bm{\rho}\cdot\bm{\rho^{\prime}}\right]}{4\sinh(\omega_{\perp}\beta)}\right\} ;\label{k1}\\
K_{\beta}^{(e)}(z_{e},z_{e}^{\prime}) & = & \sqrt{\frac{\omega_{z}}{2\pi\sinh(\omega_{z}\beta)}}\exp\left\{ -\frac{\omega_{z}\left[(z_{e}^{2}+z_{e}^{\prime2})\cosh(\omega_{z}\beta)-2z_{e}z_{e}^{\prime}\right]}{2\sinh(\omega_{z}\beta)}\right\} ;\label{k2}\\
K_{\beta}^{(g)}(z_{g},z_{g}^{\prime}) & = & \sqrt{\frac{1}{2\pi\beta}}\exp\left[-\frac{\left(z_{g}-z_{g}^{\prime}\right)^{2}}{2\beta}\right].\label{k3}
\end{eqnarray}
Substituting Eqs. (\ref{k1})-(\ref{k3}) into Eq. (\ref{kk}), we
can obtain the expression for the function $K_{\beta}({\bf 0},z_{e},z_{g};{\bf 0},z^{\prime},z')$
in Eq. (\ref{gg}):
\begin{equation}
K_{\beta}({\bf 0},z_{e},z_{g};{\bf 0},z^{\prime},z')=\frac{\omega_{\perp}}{8\pi^{2}\sinh(\omega_{\perp}\beta)}\sqrt{\frac{\omega_{z}}{\beta\sinh(\omega_{z}\beta)}}\exp\left\{ -\frac{\omega_{z}\left[\left(z_{e}^{2}+z^{\prime2}\right)\cosh(\omega_{z}\beta)-2z_{e}z^{\prime}\right]}{2\sinh(\omega_{z}\beta)}-\frac{(z_{g}-z^{\prime})^{2}}{2\beta}\right\} .\label{k02}
\end{equation}

In our following calculations we will use Eqs. (\ref{gre}), (\ref{gg})
and (\ref{k02}).

\subsection{Short-range behavior of $\Psi_{\xi}({\bf 0},z-\frac{z_{r}}{2},z+\frac{z_{r}}{2})$}

Now we derive the equation for $\eta_{\xi}(z)$. According to Eq.
(\ref{etaz}), $\eta_{\xi}(z)$ is determined by the behavior of the
function $\Psi_{\xi}({\bf 0},z-\frac{z_{r}}{2},z+\frac{z_{r}}{2})$
in the short-range limit $|z_{r}|\rightarrow0$. Thus, we first study
this behavior . According to Eq. (\ref{lse}) and Eq. (\ref{gre}),
the function $\Psi_{\xi}({\bf 0},z-\frac{z_{r}}{2},z+\frac{z_{r}}{2})$
satisfies the equation
\begin{eqnarray}
 &  & \Psi_{\xi}({\bf 0},z-\frac{z_{r}}{2},z+\frac{z_{r}}{2})=\Psi^{(0)}({\bf 0},z-\frac{z_{r}}{2},z+\frac{z_{r}}{2})+2\omega_{\perp}a_{\xi}\int dz'g\left[E-\omega_{\perp};z-\frac{z_{r}}{2},z+\frac{z_{r}}{2};z',z'\right]\eta_{\xi}(z')\nonumber \\
 &  & +4\pi a_{\xi}\eta_{\xi}(z)\int dz'G_{E^{\prime}}({\bf 0},z-\frac{z_{r}}{2},z+\frac{z_{r}}{2};{\bf 0},z^{\prime},z')+4\pi a_{\xi}\int dz'G_{E^{\prime}}({\bf 0},z-\frac{z_{r}}{2},z+\frac{z_{r}}{2};{\bf 0},z^{\prime},z')\left[\eta_{\xi}(z')-\eta_{\xi}(z)\right],\nonumber \\
\label{lse2}
\end{eqnarray}
In the limit $z_r\rightarrow0$, the 1st and 2nd term in
the right-hand side of the above equation converges. Now we study
the behavior of the 3rd term. To this end, we define
\begin{equation}
U\equiv\int dz'G_{E^{\prime}}({\bf 0},z-\frac{z_{r}}{2},z+\frac{z_{r}}{2};{\bf 0},z^{\prime},z').\label{ud}
\end{equation}
Then Eq. (\ref{gg}) yields that
\begin{eqnarray}
U & = & \int_{0}^{\infty}d\beta{\cal L}\left[\beta,z,z_{r}\right],\label{uu}
\end{eqnarray}
with the function ${\cal L}\left[\beta,z,z_{r}\right]$ being defined
as
\begin{equation}
{\cal L}[\beta,z,z_{r}]=-\int_{-\infty}^{\infty}dz^{\prime}e^{\beta E^{\prime}}K_{\beta}\left({\bf 0},z-\frac{z_{r}}{2},z+\frac{z_{r}}{2};{\bf 0},z^{\prime},z'\right).\label{ll}
\end{equation}
Substituting Eq. (\ref{k02}) into Eq. (\ref{ll}), we further obtain
\begin{eqnarray}
{\cal L}[\beta,z,z_{r}] & = & -\frac{\omega_{\perp}\sqrt{\omega_{z}}}{2(2\pi)^{\frac{3}{2}}\sinh(\omega_{\perp}\beta)\sqrt{\omega_{z}\beta\cosh(\omega_{z}\beta)+\sinh(\omega_{z}\beta)}}\nonumber \\
 &  & \times\exp\left\{ \beta E^{\prime}+\frac{\left[\frac{\omega_{z}(z-\frac{z_{r}}{2})}{\sinh(\omega_{z}\beta)}+\frac{(z+\frac{z_{r}}{2})}{\beta}\right]^{2}}{2\left(\frac{\omega_{z}}{\tanh(\omega_{z}\beta)}+\frac{1}{\beta}\right)}-\frac{\omega_{z}(z-\frac{z_{r}}{2})^{2}}{2\tanh(\omega_{z}\beta)}-\frac{1}{2\beta}\left(z+\frac{z_{r}}{2}\right)^{2}\right\} .\label{ll2}
\end{eqnarray}
Furthermore, the integration $\int_{0}^{\infty}d\beta{\cal L}[\beta,z,z_{r}]$
diverges in the limit $z_{r}\rightarrow0$. This divergence is due
to the behavior of ${\cal L}[\beta,z,z_{r}]$ in the limit $\beta\rightarrow0$.
Thus, we can obtain the behavior of $\int_{0}^{\infty}d\beta{\cal L}[\beta,z,z_{r}]$
in this limit by re-expressing ${\cal L}[\beta,z,z_{r}]$ as
\begin{equation}
{\cal L}[\beta,z,z_{r}]={\cal L}_{0}[\beta,z,z_{r}]+{\cal L}_{1}[\beta,z,z_{r}],\label{ll3}
\end{equation}
where
\[
{\cal L}_{0}[\beta,z,z_{r}]={\cal L}[\beta\rightarrow0^{+},z,z_{r}]=\frac{1}{8(\pi\beta)^{3/2}}\exp\left(-\frac{z_{r}^{2}}{4\beta}\right),
\]
and
\[
{\cal L}_{1}[\beta,z,z_{r}]={\cal L}[\beta,z,z_{r}]-{\cal L}_{0}[\beta,z,z_{r}].
\]
Thus, we have 
\begin{align}
U=\int_{0}^{\infty}{\cal L}_{0}[\beta,z,z_{r}]d\beta+\int_{0}^{\infty}{\cal L}_{1}[\beta,z,z_r]=-\frac{1}{4\pi|z_{r}|}+F_{1}(z)+{\cal O}(z_{r})\label{uexp}
\end{align}
where the function $F_{1}(z)$ is given by
\begin{eqnarray}
F_{1}(z) & = & \int_{0}^{\infty}{\cal L}_{1}[\beta,z,z_{r}=0]d\beta\nonumber \\
 & = & \int_{0}^{\infty}\left\{ {\cal L}[\beta,z,z_{r}=0]-{\cal L}_{0}[\beta,z,z_{r}=0]\right\} d\beta\nonumber \\
 & = & -\frac{1}{4\pi^{\frac{3}{2}}}\int_{0}^{\infty}d\beta\left[\frac{\omega_{\perp}\sqrt{\omega_{z}}\exp\left(\beta E^{\prime}-\frac{\omega_{z}\left[\omega_{z}\beta+2\tanh(\omega_{z}\beta/2)\right]}{2[1+\omega_{z}\hbar\beta\coth(\omega_{z}\beta)]}z^{2}\right)}{\sqrt{2}\sinh(\beta\omega_{\perp})\sqrt{\omega_{z}\beta\cosh(\omega_{z}\beta)+\sinh(\omega_{z}\beta)}}-\frac{1}{2\beta^{3/2}}\right].\label{f1}
\end{eqnarray}
Using the result in Eqs. (\ref{uu}, \ref{uexp}) we can obtain the
behavior of the 3rd term in the right-hand-side of Eq. (\ref{lse2})
in the limit $|z_{r}|\rightarrow0$.

Finally, we can show that last term in the right-hand-side of Eq.
(\ref{lse2}) is convergent in the limit $|z_{r}|\rightarrow0$ with
the following analysis, which is quite similar to the analysis around
Eq. (E12) of Ref. \cite{castin}. This term is proportional to $\int_{-\infty}^{+\infty}dz'G_{E^{\prime}}({\bf 0},z-\frac{z_{r}}{2},z+\frac{z_{r}}{2};{\bf 0},z^{\prime},z^{\prime})\left[\eta_{\xi}(z')-\eta_{\xi}(z)\right]$.
By definding $u=z^{\prime}-z$, we can re-write this integration as
$I\equiv\int_{0}^{+\infty}du\left[{\cal G}(z_{r};u)B(u)+{\cal G}(z_{r};-u)B(-u)\right]$,
with ${\cal G}(z_{r};u)=G_{E^{\prime}}({\bf 0},z-\frac{z_{r}}{2},z+\frac{z_{r}}{2};{\bf 0},z+u,z+u)$
and $B(u)=\eta_{\xi}(z+u)-\eta_{\xi}(z)$. In the limit $|z_{r}|\rightarrow0$,
the only possible cause for the divergence of $I$ is the fact that
${\cal G}(z_{r}=0;u)$ diverges as $1/u^{2}$ for $u\rightarrow0$.
However, when $u\rightarrow0$ we also we have $B(\pm u)=\pm B^{\prime}u+B^{\prime\prime}u^{2}$,
with $B^{\prime}=\left.dB(u)/du\right|_{u=0}$ and $B^{\prime\prime}=\left.d^{2}B(u)/du^{2}\right|_{u=0}$,
which leads to $[{\cal G}(z_{r};u)B(u)+{\cal G}(z_{r};-u)B(-u)]\propto\frac{1}{u^{2}}[2B^{\prime\prime}u^{2}+{\cal O}(u^{3})]\propto2B^{\prime\prime}+{\cal O}(u)$.
Notice that the linear terms $\pm B^{\prime}u$ in $B(u)$ and $B(-u)$
cancel with each other. Thus, the divergence of ${\cal G}(z_{r}=0;u)$
is canceled by the functions $B(\pm u)$, and the integration
$I$ and the last term in the right-hand-side of Eq. (\ref{lse2})
is thus convergent.

With our above results, especially Eqs. (\ref{uu}, \ref{uexp}),
we obtain the behavior of $\Psi_{\xi}({\bf 0},z-\frac{z_{r}}{2},z+\frac{z_{r}}{2})$
in the short-range limit $|z_{r}|\rightarrow0$:

\begin{eqnarray}
\lim_{z_{r}\rightarrow0}\Psi_{\xi}({\bf 0},z-\frac{z_{r}}{2},z+\frac{z_{r}}{2}) & = & -\frac{1}{|z_{r}|}a_{\xi}\eta_{\xi}(z)+\Psi^{(0)}({\bf 0},z,z)+2\omega_{\perp}a_{\xi}\int dz'g\left[E-\omega_{\perp};z,z;z',z'\right]\eta_{\xi}(z')\nonumber \\
 &  & +4\pi a_{\xi}F_{1}(z)\eta_{\xi}(z)+4\pi a_{\xi}\int dz'F_{2}(z,z^{\prime})\left[\eta_{\xi}(z')-\eta_{\xi}(z)\right]+{\cal O}(z_{r}),\label{psi3}
\end{eqnarray}
where the function $F_{2}(z,z^{\prime})$ is defined as $F_{2}(z,z^{\prime})\equiv G_{E^{\prime}}({\bf 0},z,z;{\bf 0},z^{\prime},z').$
Using Eq. (\ref{gg}) and Eq. (\ref{k02}), we obtain
\begin{equation}
F_{2}(z,z^{\prime})=-\int_{0}^{\infty}d\beta\frac{\omega_{\perp}}{8\pi^{2}\sinh(\omega_{\perp}\beta)}\sqrt{\frac{\omega_{z}}{\beta\sinh(\omega_{z}\beta)}}\exp\left\{ \beta E^{\prime}-\frac{\omega_{z}\left[\left(z^{2}+z^{\prime2}\right)\cosh(\omega_{z}\beta)-2zz^{\prime}\right]}{2\sinh(\omega_{z}\beta)}-\frac{(z-z^{\prime})^{2}}{2\beta}\right\} .\label{f2}
\end{equation}

\subsection{Integral equaiton for $\eta_{\xi}(z)$}

Substituting Eq. (\ref{f2}) into Eq. (\ref{etaz}), we obtain the
integral equation for $\eta_{\xi}(z)$:
\begin{equation}
\eta_{\xi}(z)=\Psi^{(0)}\left({\bf 0},z,z\right)+\hat{O}_{\xi}[\eta_{\xi}(z)],\label{eta3}
\end{equation}
where $\hat{O}_{\xi}$ is an integral operator which is defined as
\begin{eqnarray}
\hat{O}_{\xi}[\eta_{\xi}(z)] & \equiv & 2\omega_{\perp}a_{\xi}\int dz'g\left[E-\omega_{\perp};z,z;z',z'\right]\eta_{\xi}(z')+4\pi a_{\xi}F_{1}(z)\eta_{\xi}(z)+4\pi a_{\xi}\int dz'F_{2}(z,z^{\prime})\left[\eta_{\xi}(z')-\eta_{\xi}(z)\right],\nonumber \\
\label{eta33}
\end{eqnarray}
with the function $g\left[E-\omega_{\perp};z,z;z',z'\right]$ being
defined in Eq. (\ref{a2}), the function $F_{1}(z)$ being defined
in Eq. (\ref{f1}) and the function $F_{2}(z,z^{\prime})$ being defined
in Eq. (\ref{f2}). Eq. (\ref{eta3}) is just Eq. (\ref{eta2}) in
our main text.

\end{widetext}


\begin{thebibliography}{1}



\bibitem{hubbar1}
S. Sugawa, K. Inaba, S. Taie,    R. Yamazaki, M. Yamashita and Y. Takahashi, Nat. Phys. {\bf 7}, 642 (2011).

\bibitem{hubbard2}
S. Taie, R. Yamazaki, S. Sugawa    and Y. Takahashi, Nat. Phys. {\bf 8}, 825 (2012).

\bibitem{ca} G. Wilpers, T. Binnewies, C. Degenhardt, U. Sterr, J. Helmcke, and F. Riehle, Phys. Rev. Lett. {\bf 89} 230801 (2002)

\bibitem{hubbard3}
C. Hofrichter, L. Riegger, F. Scazza, M. H\"ofer, D. R. Fernandes, I. Bloch, and S. F\"olling, Phys. Rev. X {\bf 6}, 021030 (2016).

\bibitem{martin}
M. J. Martin, M. Bishof, M. D. Swallows, X. Zhang, C. Benko, J. von-Stecher, A. V. Gorshkov, A. M. Rey and Jun Ye, Science {\bf 341}, 632 (2013).

\bibitem{Jun}
X. Zhang, M. Bishof, S. L. Bromley, C. V. Kraus, M. S. Safronova, P. Zoller, A. M. Rey, J. Ye, Science {\bf 345}, 1467 (2014).

\bibitem{1dsun}
G. Pagano, M. Mancini, G. Cappellini, P. Lombardi, F. Sch\"afer, H. Hu, X.-J. Liu, J. Catani, C. Sias, M. Inguscio and L. Fallani, Nat. Phys. {\bf 10}, 198 (2014).

\bibitem{review} M. A. Cazalilla and A. M. Rey, Rep. Prog. Phys. {\bf 77}, 124401 (2014).

\bibitem{Munich}
F. Scazza, C. Hofrichter, M. H\"ofer, P. C. De Groot, I. Bloch, and S. F\"olling, Nat. Phys. {\bf 10}, 779 (2014) and Nat. Phys. {\bf 11}, 514 (2015).

\bibitem{Florence}
G. Cappellini, M. Mancini, G. Pagano, P. Lombardi, L. Livi, M. Siciliani de Cumis, P. Cancio, M. Pizzocaro, D. Calonico, F. Levi, C. Sias, J. Catani, M. Inguscio, and L. Fallani, Phys. Rev. Lett. {\bf 113}, 120402 (2014) and Phys. Rev. Lett. {\bf 114}, 239903 (2015).

\bibitem{ourprl}
R. Zhang, Y. Cheng, H. Zhai, and P. Zhang, Phys. Rev. Lett. {\bf 115}, 135301 (2015).

\bibitem{ofrexp1}
G. Pagano, M. Mancini, G. Cappellini, L. Livi, C. Sias, J. Catani, M. Inguscio, L. Fallani, Phys. Rev. Lett. {\bf 115}, 265301 (2015).

\bibitem{ofrexp2}
M. H\"ofer, L. Riegger, F. Scazza, C. Hofrichter, D.R. Fernandes, M. M. Parish, J. Levinsen, I. Bloch, S. F\"olling, Phys. Rev. Lett. {\bf 115}, 265302 (2015).

\bibitem{soc1}
M. L. Wall, A. P. Koller, S. Li, X. Zhang, N. R. Cooper, J. Ye and A. M. Rey, Phys. Rev. Lett. {\bf 116}, 035301 (2016).

\bibitem{soc2}
L. F. Livi, G. Cappellini, M. Diem, L. Franchi, C. Clivati, M. Frittelli, F. Levi, D. Calonico, J. Catani, M. Inguscio, and L. Fallani, Phys. Rev. Lett. {\bf 117}, 220401 (2016).

\bibitem{soc3}
S. Kolkowitz, S. L. Bromley, T. Bothwell, M. L. Wall, G. E. Marti, A. P. Koller, X. Zhang, A. M. Rey, J. Ye, Nature {\bf 542}, 66 (2017).

\bibitem{kondo} A. C. Hewson, \textit{The Kondo Problem to Heavy Fermions}, Cambridge University Press, 1993.

\bibitem{kondo-Rey}
A. V. Gorshkov, M. Hermele, V. Gurarie, C. Xu, P. S. Julienne, J. Ye, P. Zoller, E. Demler, M. D. Lukin, and A. M. Rey, Nat. Phys. {\bf 6}, 289 (2010).

\bibitem{kondo-rey2} 
J. Bauer, C. Salomon, and E. Demler, Phys. Rev. Lett. {\bf 111}, 215304 (2013).

\bibitem{kondo-salomon} 
L. Isaev and A. M. Rey, Phys. Rev. Lett. {\bf 115}, 165302 (2015).


\bibitem{kondo-Jo}
I. Kuzmenko, T. Kuzmenko, Y. Avishai, G. B. Jo, Phys. Rev. B {\bf 93}, 115143 (2016) and Phys. Rev. B {\bf 97}, 075124 (2018)

\bibitem{our1}
R. Zhang, D. Zhang, Y. Cheng, W. Chen, P. Zhang, and H. Zhai, Phys. Rev. A {\bf 93}, 043601 (2016).

\bibitem{our2}
Y. Cheng, R. Zhang, P. Zhang, and H. Zhai, Phys. Rev. A {\bf 96}, 063605 (2017).

\bibitem{blochexp}
L. Riegger, N. D. Oppong, M. H\"ofer, D. R. Fernandes, I. Bloch and Simon F\"olling, Phys. Rev. Lett. {\bf 120}, 143601 (2018).


\bibitem{magic1} Z. W. Barber, J. E. Stalnaker, N. D. Lemke, N. Poli, C. W. Oates, T. M. Fortier, S. A. Diddams, L. Hollberg, C. W. Hoyt, A. V. Taichenachev, and V. I. Yudin, Phys. Rev. Lett. {\bf 100}, 103002 (2008).

\bibitem{magic2}V. A. Dzuba and A. Derevianko, J. Phys. B: At. Mol. Opt. Phys. {\bf 43} 074011 (2010).


\bibitem{magton}
K. Jachymski, T. Wasak, Z. Idziaszek,
P. S. Julienne, A. Negretti, and T. Calarco, Phys. Rev. Lett. {\bf 120}, 013401 (2018).

\bibitem{magton2}
T. Wasak, K. Jachymski, T. Calarco, A. Negretti, arXiv:1803.03024.


\bibitem{olshaniiodd} M. D. Girardeau1, and M. Olshanii, Phys. Rev. A {\bf 70}, 023608 (2004).




\bibitem{castin}
P. Massignan and Y. Castin, Phys. Rev. A {\bf 74}, 013616 (2006).


\bibitem{olshanii} M. Olshanii, Phys. Rev. Lett. {\bf 81}, 938 (1998).

\bibitem{fodd1} L. Pricoupenko, Phys. Rev. Lett. {\bf 100}, 170404 (2008).

\bibitem{fodd2} A. Imambekov, A. A. Lukyanov, L. I. Glazman, and V. Gritsev, Phys. Rev. Lett. {\bf 104}, 040402 (2010).

\bibitem{fodd3} L. Zhou and X. Cui, Phys. Rev. A {\bf 96}, 030701(R) (2017).

\bibitem{fodd4} X. Cui, Phys. Rev. A {\bf 94}, 043636 (2016).

\bibitem{fodd5} X. Cui and H. Dong, Phys. Rev. A {\bf 94}, 063650 (2016).

\bibitem{fodd6} X. Cui, Phys. Rev. A {\bf 95}, 041601(R) (2017).


\bibitem{tannishida} Y. Nishida and S. Tan, Phys. Rev. A {\bf 82}, 062713 (2010).



\bibitem{tan-mixd-d}
Y. Nishida and S. Tan, Phys. Rev. Lett. {\bf 101}, 170401 (2008).

\bibitem{rccouple1}
S. Sala, P. Schneider, and A. Saenz, Phys. Rev. Lett. {\bf 109}, 073201 (2012).

\bibitem{rccouple2}
V. Peano, M. Thorwart, C. Mora and R. Egger, New. Jour. Phys. {\bf 7}, 192 (2005).

\bibitem{rccouple3} G. Lamporesi, J. Catani, G. Barontini, Y. Nishida, M. Inguscio, and F. Minardi, Phys. Rev. Lett. {\bf 104}, 153202 (2010).

\bibitem{rccouple4} Y. Nishida and S. Tan, Phys. Rev. A {\bf 82}, 062713 (2010).

\bibitem{rccouple5} F Minardi, G Barontini, J Catani, G Lamporesi, Y Nishida and M Inguscio, J. Phys.: Conf. Ser. {\bf 264} 012016 (2011).

\bibitem{rccouple6} J. P. Kestner and L-M. Duan, New. Jour. Phys, {\bf 12},  053016 (2010).

\bibitem{rccouple7} S.-G. Peng, H. Hu, X.-J. Liu, and P. D. Drummond,  Phys. Rev.
A {\bf 84}, 043619 (2011).


\end{thebibliography}
\end{document}